\RequirePackage{fix-cm}
\documentclass[smallextended]{svjour3}       
\smartqed  
\usepackage{graphicx}
\usepackage{amsmath}
\usepackage{natbib}
\usepackage{booktabs}
\usepackage{xcolor}
\usepackage{longtable}
\usepackage{url}
\usepackage{placeins}

\usepackage{tcolorbox}
\definecolor{light-gray}{gray}{0.85}
\newcommand{\summaryboxnew}[1]{
\begin{tcolorbox}[colback=light-gray!5!light-gray,colframe=light-gray!75!light-gray,top=2mm,left=2mm,bottom=2mm,right=2mm]
  #1
\end{tcolorbox}
}
\begin{document}

\newcommand{\rev}[1]{\textcolor{black}{#1}}

\title{The whos, whats, and whys of issues related to personal data and data protection in open-source projects on GitHub
\thanks{This research is supported by funding from the project ``Engineering Secure Systems'', topic 46.23.01 Methods for Engineering Secure Systems, of the Helmholtz Association (HGF), and by KASTEL Security Research Lab.}
}

\titlerunning{The whos, whats, and whys of issues related to data protection on GitHub}        

	\author{
        Anne Hennig \and
        Lukas Schulte \and
        Steffen Herbold \and
        Oksana Kulyk \and
        Peter Mayer
	}
	
	\institute{
        Anne Hennig, 0000-0002-6964-589X \at
		Karlsruhe Institute of Technology, Karlsruhe, Germany.
		\email{anne.hennig@kit.edu}  
		\and
        Lukas Schulte, 0000-0001-9336-2075 \at
		University of Passau, Passau, Germany.
		\email{lukas.schulte@uni-passau.de}  
		\and 
		Steffen Herbold, 0000-0001-9765-2803 \at
		University of Passau, Passau, Germany.
		\email{steffen.herbold@uni-passau.de} 
		\and 
        Oksana Kulyk, 0000-0003-4218-1658 \at
		IT University of Copenhagen, Copenhagen, Denmark.
		\email{okku@itu.dk} 
		\and 
		Peter Mayer, 0000-0002-6267-4874 \at
		University of Southern Denmark, Odense, Denmark, \&
        Karlsruhe Institute of Technology, Karlsruhe, Germany.
		\email{mayer@imada.sdu.dk}   
	}

\date{Received: date / Accepted: date}

\maketitle

\begin{abstract}
Data protection regulations such as \rev{the General Data Protection Regulation (GDPR) in the European Union} and \rev{the California Consumer Privacy Act (CCPA) in the US} affect how software may handle the personal data of its users. Prior literature focused on how \rev{data protection regulations are discussed} for software in operation\rev{,} or how this \rev{topic} is discussed in various channels outside of the software development process. Yet, what is missing, is a perspective on the impact of such regulations on the software development process. In our work, we address this gap\rev{,} and explore how discussions during the \rev{development of }software are impacted by regulations, who reports and discusses issues related to personal data and data protection, and how developers react \rev{to those issues}. To that end, we used inductive coding to analyze 652 issues from Open Source GitHub projects and used the codes to quantitatively analyze the relation between the roles, resolutions, and data protection issues to understand correlations and predict resolutions of issues. Most notably we observed a significant increase in reporting when GDPR came into effect. The most common issue types were feature requests for privacy enhancement, which were mainly reported and discussed by frequent reporters and frequent committers. But especially issues regarding privacy enhancement were also frequently reported by one-time reporters. Most of the requests were solved without opposing votes. All in all, our findings indicate that data protection regulations effectively start discussions about privacy within the software development community.
    
\end{abstract}

\section{Introduction}
\label{intro}
Multiple governments across the globe have enacted stricter data protection laws in recent years. The most notable examples are the EU's ePrivacy Directive (ePD), which was passed in 2002 and amended in 2009~\citep{ePD}, the General Data Protection Regulation (GDPR) by the European Union (EU), which came into effect in May 2018~\citep{GDPR}, and the California Consumer Privacy Act (CCPA) of 2018 \citep{CCPA}. At the core of these regulations are rules that strengthen data protection and force businesses to require valid consent by users for collecting and processing their personal data. The most visible example are cookie banners. Since these rules are fairly new, their interpretation and implementation is still evolving. When, e.g., the ePrivacy Directive was initially enacted, many websites only informed users that they collect data using cookies without allowing options to object to that data collection.

With the European Court of Justice's (ECJ) ``Planet49'' decision (judgment of 1.10.2019 - C-673/17) it became clear that consent for data collection is not freely given, affirmative, informed, and, thus, valid if an opt-out design is used\rev{. Which is, e.g.,} pre-ticked boxes on an online subscription form or – in the case of cookie disclaimers – pre-selected ``Agree'' options, which users have to actively deselect to refuse consent \citep{planet49}. Further regulations that had an effect on the design of cookie banners were given in the ``Orange România'' decision of the ECJ (judgment of 11.11.2020 - C-61/19). It was stated that the free decision of users is disproportionately constrained if the refusal of consent represents a greater effort than the granting of consent. This meant that, e.g., cookie banners where dissent options are hidden in a text or less visible than consent options, can be considered non-compliant with the GDPR \citep{orangeromania}. This is just one example of how developers, in particular those who offer or maintain a website, needed to constantly keep up with changes due to new \rev{data protection} regulations over the years.

There is already literature on how software development fails to comply with regulations \citep[e.g.,][]{Andow.2020, Chitkara.2017, Habib.2020, Matte.2020, Kampanos.2021, Bollinger.2022, Trevisan.2019} and how developers discuss privacy related topics \citep[e.g.,][]{Greene.2018, Tahaei.2020, Li.2018}. These studies have in common that they consider the impact of data protection regulations on software \textit{in operation}\rev{,} or consider data protection discussions \textit{in general}. What \rev{is} still missing\rev{,} is a perspective of how data protection regulations affect the software \textit{development} itself, \rev{and how the implementation of changes that come with new regulations are discussed}. Software products may need to be constantly updated to provide new features that enable compliance with data protection laws, e.g., in the case of cookie banners. 

In order to fill this gap in the literature, we want to shed light on the impact of data protection regulations like the GDPR on software development on GitHub. 
Our focus is on the reporting of issues related to personal data and data protection, e.g., change requests, questions, or problems with respect to data protection. In analyzing entire discussions, we also aim to understand, if reported issues are implemented or - if not - why they are not implemented. To the best of our knowledge, this has not been studied before. In summary, we want to understand \textit{how often} \rev{data protection} topics are reported, \textit{what} \rev{issues are} reported, \textit{who} reports data protection issues, and what \textit{the reaction} of developers issues is. We used qualitative and quantitative methods to analyze overall 652 issues \rev{from Open Source projects on GitHub}. Based on this data, the contributions of our research are the following: 

\begin{itemize}
  \item We found that the majority of issues are reported by developers who are frequently active within the project, either by only reporting issues (frequent reporter) or by code contributions (frequent committers). However, especially for feature requests that would enhance users' privacy or the reporting of existing features with implication for users' privacy, we also observed many one-time reporters, only active within a single issue. But while these are active as reporters, they only play a minor role in the discussions of the issues they reported.
  \item The discussed issues are of various types, with the majority of issues asking for features to enhance users' privacy. Most issues were addressed as requested, either by implementing a feature, solving the bug or providing documentation. However, beyond these common cases, there is a large diversity of other issues \rev{related to personal data and data protection} being reported that \rev{we identified in our analysis}. 
  \item We identified GDPR as a key driver for the reporting of data privacy issues. Other legislation like CCPA seems to play only a minor role.
\end{itemize}

The remainder of this paper is structured as follows: In Section~\ref{relatedwork}, we give an overview of the previous work that motivated our research questions. Then, we describe our methodology in Section~\ref{methodology}. In Section~\ref{results}, we describe the results and discuss our results in Section~\ref{discussion}. We conclude the paper in Section~\ref{summary}.

\section{Related Work}
\label{relatedwork}

Due to its huge number of users and projects, GitHub is a common source for historic data about software development, e.g., when studying social aspects of software development~\citep{Herbold2021}, code reviews~\citep{Rao2022}, or source code history~\citep{Herbold2022}. There is also research particular to GitHub issues, e.g., regarding their labels~\citep{xie2021mula} and types~\citep{Kallis2022}. Furthermore, there is prior work that focuses on particular aspects of discussions on GitHub, e.g., security~\citep{Pletea.2014}, end-user issues~\citep{Khalajzadeh.2022}, or the relation of emotion and issue closing~\citep{Ferreira2022}. While not directly related to our work, these papers demonstrate the suitability of GitHub and GitHub issues as study subjects, including \rev{identifying} how developers deal with specific aspects of software, such as its security. 

Previous research has \rev{also} shown that despite the introduction of stricter data protection laws in recent years, software still fails to comply to these regulations \citep[e.g.,][]{Andow.2020, Chitkara.2017, Habib.2020, Matte.2020, Kampanos.2021, Bollinger.2022, Trevisan.2019}. Therefore, the question arises how data protection topics are discussed among developers \rev{and what hinders developers to fully comply to data protection regulations}. There is, e.g., research analyzing discussions on privacy-related \rev{topics} among developers via lab studies~\citep{Senarath.2018}, in-depth interviews~\citep{Li.2018, Hadar.2018, Kekulluoglu.2023, Peixoto.2023, Iwaya.2023} or online surveys~\citep{Sheth.2014, Balebako.2014, Kuetreiber.2022}.

While lab studies, interviews and surveys provide important insights into how developers understand privacy and data protection in general, these methods initially stimulated the participants to think about \rev{topics} related to data protection. \rev{Instead,} analyses of discussions in various forums are helpful to gain more genuine opinions or understandings without giving a prompt~\citep{Li.2021}.\footnote{Forums have been used in various contexts to analyze topics that are discussed among developers, e.g. on Stack~Overflow~\citep{Abdalkareem.2017, Wu.2019, Barua.2014}, or as a comparison between Stack~Overflow and GitHub~\citep{Vasilescu.2013, Han.2020}.} Most notably, three analyses were conducted on understanding personal data and privacy discussions in developer forums~\citep{Li.2021, Tahaei.2020, Greene.2018}.

\cite{Greene.2018} analyzed privacy-related discussion in an iOS developer forum (iPhoneDevSDK) and an Android forum (XDA) to identify how the term ``privacy'' is discussed, defined, and framed in contrasting communities. The authors found that privacy is highly influenced by the platform's philosophy and, therefore, used fundamentally different in both forums. The authors suggest that for the mobile context, the discussion rather followed a ``privacy by platform'' than a ``privacy by design'' approach.

\cite{Tahaei.2020} used topic modeling techniques to identify privacy-related questions on Stack~Overflow and qualitatively analyzed a random sample. The authors identified the following topics which are discussed among developers: privacy policies, privacy concerns, access control, and version changes, with app-related questions being an overarching issue for developers. The authors found that personal concerns as well as client or company requirements were mainly inspiring discussions, whereas laws and regulations, e.g., GDPR, were the least common drivers for discussions. This is also reflected in the work of \cite{Bissyande.2013}, where ``privacy'' was not among the top 10 topics reported in a sample of around 20,000 open source projects on GitHub. 

\cite{Li.2021} conducted a qualitative analysis of Reddit posts on issues related to personal data in an Android developer forum. Interestingly, the authors found that most developers rarely discussed privacy concerns during the development or implementation of an app, but rather when the discussion was stimulated by external events, e.g., new privacy laws.

However, all of these analyses are focused mainly on other platforms, like Stack~Overflow, and/or the mobile context. To the best of our knowledge, there is no work directly related to ours, i.e., studying issue discussions about personal data and data protection on GitHub for Open Source Software in general. Considering this open question, we state our first research question:

\begin{description}
  \item \textbf{RQ 1: \textit{What kind of issues related to personal data and data protection are discussed on GitHub?}}
\end{description}

\rev{By investigating this question, our study is the first one to analyze how data protection and personal data is discussed during the software development process, what aspects of personal data and data protection are discussed, and what stimulated the discussion. By answering this research question, we provide valuable information which impact, e.g., data protection regulations have on the software development process. Furthermore, we provide first insights into which aspects of personal data and data protection -- and which not -- are discussed among developers.}

Interestingly, \cite{Tahaei.2020} found that the first questions on ``privacy'' were created in 2008 (when Stack~Overflow was launched), followed by a – more or less – continuous increase in questions over the next 10 years. 
There was no evaluation of whether the reasons driving these discussions or the topics have changed over the years. Therefore, it would be interesting to see, if, e.g., an increase in discussions driven by data protection law correlates with amendments to the law. We want to address this gap with our second research question: 

\begin{description}
    \item \textbf{RQ 2: \textit{How often are those issues related to personal data and data protection reported?}}
\end{description}

\rev{By answering this research question, we provide quantitative data whether and when data protection regulations have an impact on the software development process. This information might also prove the effectiveness of such regulations, and shed light on whether and when in the software development process regulations stimulate developers to think about data protection and protection of personal data.}

Another open question is, how different individuals are interacting in data protection-related discussions and what this means for the software development process. \cite{Bissyande.2013} characterized the reporting behavior for open source projects on GitHub in general. The authors found that the majority of projects record a small amount of reported issues. Only about 8\% of projects were found to have more than 100 issues reported. Most issues are reported by developers with a large number of followers, and mainly for larger and established projects with popular owners and a large number of watchers and forks. Furthermore, the authors found that issue reporters, even if they do not belong to the development team, contribute to the code base in most cases. 

Besides the number of reporters and their contribution to the project, there was no further classification of roles. It will be interesting to see if these findings also hold especially for issues related to data protection and personal data, or if we find, e.g., more one-time reporters who report the same issue to several software projects.

Thus, we state our third research question:
\begin{description}
    \item \textbf{RQ 3: \textit{Who reports and discusses issues related to personal data and data protection on GitHub?}}
\end{description}

\rev{By answering this research question, we can complement and extend related work on the reporting behavior on GitHub. Furthermore, it will be interesting to compare whether data protection and personal data topics attract different types of reporters and discussants, and whether these topics might motivate outsiders to contribute to the code.}

Finally, we want to shed some light on the developers' reactions. As shown in a lab study by~\cite{Senarath.2018}, developers seemed to find it difficult, to embed privacy requirements into applications. Participants complained that privacy requirements contradicted system requirements, and they had difficulties implementing privacy requirements with appropriate privacy techniques, probably due to a lack of knowledge about privacy-preserving techniques~\citep{Senarath.2018}. Previous interview studies and surveys with software developers also showed that a lack of resources~\citep{Balebako.2014} or a lack of sense of responsibility~\citep{Kuetreiber.2022}, a lack of knowledge about privacy best practices or privacy-preserving technologies~\citep{Balebako.2014, Li.2018, Hadar.2018, Kuetreiber.2022, Iwaya.2023}, and insufficient understanding of data protection concepts~\citep{Li.2018, Hadar.2018} are major issues in software development. Furthermore, \cite{Sheth.2014} found that expectations and needs regarding privacy differ between users and developers of software, leading to developers being less concerned about processing personal data and more trustworthy towards their systems than users. On the other hand, recent studies showed that software developers are increasingly aware of information privacy, privacy regulations, and privacy strategies~\citep{Iwaya.2023}, deem privacy desirable~\citep{Sheth.2014, Kekulluoglu.2023}, and have rather practical than theoretical knowledge of privacy~\citep{Peixoto.2023}. Nevertheless, this knowledge is rarely implemented~\citep{Iwaya.2023} and privacy issues are considered less important compared to security issues~\citep{Balebako.2014, Hadar.2018, Peixoto.2023, Kekulluoglu.2023}.

We, therefore, found it interesting to investigate if and how the developers handle reported issues related to personal data and data protection, and, e.g., whether or not developers share users' concerns. This motivated our fourth research question: 
\begin{description}
    \item \textbf{RQ 4: \textit{How do developers react to such reported issues?}}
\end{description}

\rev{Answering this research question in relation to the issue type and the reporter role will provide a deeper understanding of how developers react to data protection and personal data issues in an unprompted setting, i.e., neither in an interview nor a survey, where interviewer and survey effects might bias the answers. With our research, we aim to identify patterns and correlations that can be investigated in future studies in more detail.}

Thus, we study the topic from three perspectives. The first perspective is the type of issues that were reported (e.g., requests to provide additional features to enable data protection, requests for the removal of functionality to preserve data protection, or bugs that have an impact on the users' privacy). The second perspective considers the amount of reported issues and the reporters of the issues. We want to research if we see an increase in topics related to personal data and data protection just before privacy legislation came into effect. Furthermore, we want to research if the reporters of those issues come from outside the development team (e.g., end-users of software or developers who imported a library) or from inside the development team. The third perspective considers what happens after the reporting, e.g., whether issues were ignored, discussed without resolution, resolved, or rejected, and what the reason for, e.g., rejecting requests was. 
\rev{Due to the nature of this explorative research, our main goal is to describe the patterns we identify. Future research can build on theses findings to investigate certain aspects in more details and provide fine-granular explanations.}
We will describe our methodology in more detail in the next section (Section \ref{methodology}).

\section{Methodology}
\label{methodology}

\begin{figure*}
\includegraphics[width=\linewidth]{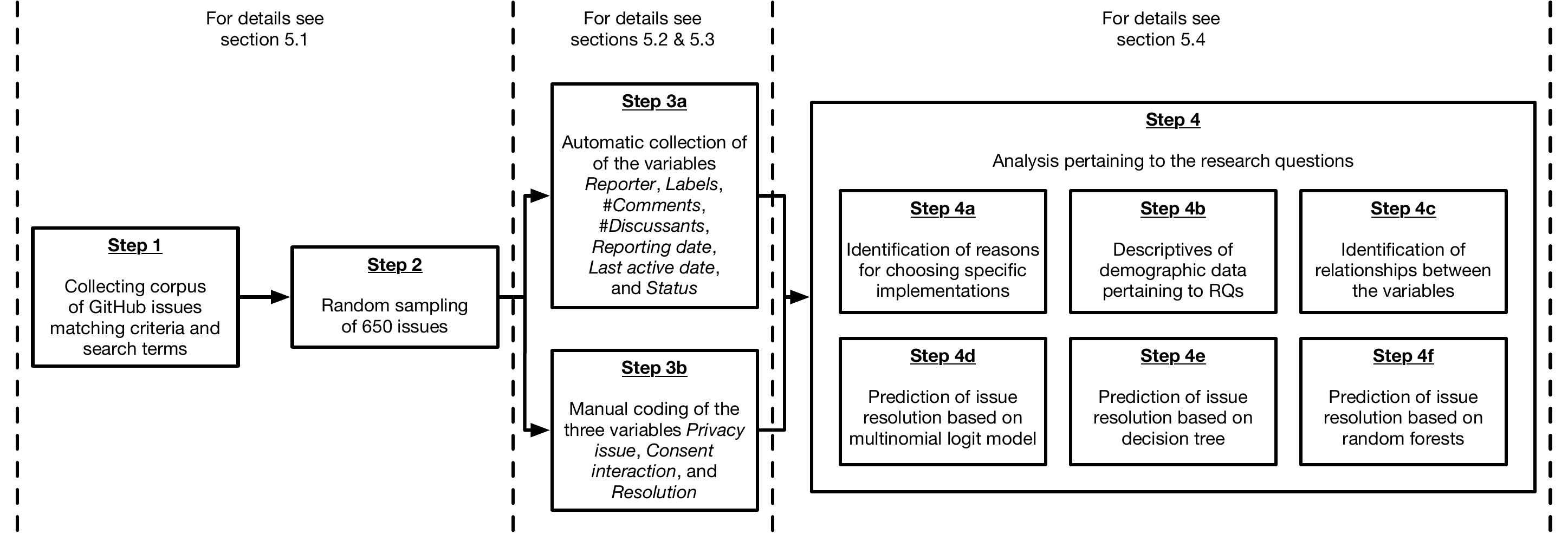}
\caption{Overview of the steps involved in the methodology of this investigation.}
\label{fig:methodology}
\end{figure*}

In the following, we define the materials in Section \ref{sec:subjects}, our variables in Section \ref{sec:variables}, and our data collection and coding in Section \ref{sec:execution-plan}. We then present the quantitative and qualitative analyses that we conducted to answer our research questions in Section \ref{sec:analysis-plan}. The study protocol was peer-reviewed \rev{and pre-registered with continuity acceptance for submission to Empirical Software Engineering granted by RR-Committee of the MSR'23}~\citep{hennig2023understanding}. An overview of all steps described in this section can be found in Figure \ref{fig:methodology}. All deviations from the registered protocol are summarized in Section~\ref{sec:deviations}.

\subsection{Materials and Subjects}
\label{sec:subjects}

Our study is based on GitHub data from April 2016\footnote{The GDRP was adopted April 14th, 2016. Other legislation like the CCPA is younger and, therefore, also covered.} until December 2022. Subsequent analysis of the reactions of developers on issues may also involve additional materials with resources from the projects we study, such as commits where an issue was addressed, mailing lists and externally hosted issue trackers, like Jira or project policies.

The subjects of our investigation are GitHub issues that mention or discuss personal data or data protection concerns, and are reported in English. {\rev{There are a lot of discussions on how to define ``personal data'' (e.g., \cite{Saglam.2022,Rupp.2024}). However, so far no conclusive definition has been found and a lot of gray areas exist. We, therefore, did not pre-define this term and left it open to the inductive coding to see, which data developers define as personal data that for them are worth protecting by data protection regulations.}} \rev{W}e, also, do not consider all of GitHub, but rather limit our investigation to issues from projects that fulfill the following criteria: 
\begin{itemize}
  \item public projects that are not forks and contain a license agreement;
  \item projects with at least ten contributors after June 2018;\footnote{The GDRP was enforceable starting May 25h, 2018. Other legislation like the CCPA is younger, and there may have been less activity after its adoption.}
  \item projects with active development that have at least 100 commits after June 2018; and
  \item active usage of GitHub issues with at least 20 issues reported after June 2018.
\end{itemize}

These criteria guarantee that the projects we consider have at least a small community, active development, and are actively using the reporting mechanism we study, i.e., GitHub issues. 

We use an automated keyword search through the GitHub search API\footnote{\url{https://docs.github.com/en/rest/search/search?apiVersion=2022-11-28\#search-issues-and-pull-requests} } to collect the data and retrieve issues for which the content contains at least one of the terms listed in Table~\ref{tbl:keywords}.
We selected these terms to cover a wide range of terminology associated with personal data and data protection, based on both the goals and aspects of data protection regulations (protection of personal data) as well as their impact (e.g., on cookie usage). We avoided terms like ``vulnerability'', ``encryption'', or ``social networks'' because they likely lead to many false positives, even though they are sometimes related to data protection issues. A search \rev{conducted between Jan 23rd and Jan 31st, 2023,} found 21,608 unique issues from 5,892 projects that meet our terms and fulfill the inclusion criteria.

We then created a random sample of 652 issues for our subsequent analysis. We manually validate for all issues in this sample if they are indeed about personal data and data protection. False positives were removed and we sampled additional issues until we reached our desired sample size. The rationale for a minimal sample size of 650 is explained when discussing our methods within the analysis (see Section~\ref{sec:analysis-plan}). Since we always sampled more issues than 650 total to reduce the number of resampling steps, we ended up with two more issues than required, i.e., 652 instead of 650. 

\begin{table}
\centering
\begin{tabular}{p{11.4cm}}
\toprule
\textbf{Terms used to find issues} \\
\midrule
anonymization,
CCPA,
consent withdrawal,
cookie banner,
cookie law,
cookie notice,
cookie prompt,
data breach,
data privacy,
data protection,
data sharing,
ePrivacy Directive,
fingerprinting,
GDPR,
personal data,
personally identifiable information,
PII,
privacy act,
privacy breach,
privacy controls,
privacy issue,
privacy law,
privacy notice,
privacy policy,
privacy problem,
privacy settings,
privacy violation,
pseudonymization,
right to be forgotten,
tracking \\ 
\bottomrule
\end{tabular}
\caption{Alphabetical list of the keywords used to identify candidates of issues regarding personal data and data protection.}
\label{tbl:keywords}
\end{table}

\subsection{Variables}
\label{sec:variables}

We measure the following variables for each of our subjects. \rev{The classification into the roles of reporter and discussants is similar to the work by, e.g., \cite{joblin2017classifying} and \cite{Honsel2016}, who differentiate between core and peripheral developers based on the activity within the project. In our work, we use a simpler approach that uses a very strict definition of peripheral, i.e., only being active once. The intent of this is to identify if somebody was active just because of one data protection issue.}

\begin{itemize}
  \item \textit{Reporter}: the role of the reporter\rev{, i.e., the person who created an issue,} within the project as one of the following: frequent reporter, one-time reporter (only active within a single issue), frequent committer, one-time committer (only active within a single commit or pull request).
  \item \textit{Discussants}: the role of the discussants\rev{, i.e., the people involved in the discussion of an issue} within the project, analogue to the roles of the reporters \rev{(frequent reporter, one-time reporter, frequent committer, one-time committer)}. 
  \item \textit{Labels}: the labels assigned to the issue on GitHub, e.g., bug, question, or enhancement. In order to avoid multiple labels with different names, but the same semantics (e.g., bug and defect), we manually generated a mapping for synonyms similar to the work by \cite{Herbold2022}, where this was done for Jira issue types (see Appendix \ref{sec:label-map}).
  \item \textit{\#Comments}: the number of comments in the discussion of the issue.
  \item \textit{\#Discussants}: the number of individuals involved in the discussion.
  \item \textit{Reporting date}: the date the issue was reported.
  \item \textit{Trigger}: the reason that triggered the reporting of the issue. 
  \item \textit{Privacy issue}: type of privacy issue.\footnote{While we use only the term privacy here, this includes personal data and data protection related issues as well.}
  \item \textit{Consent interaction}: interaction to obtain data collection consent. 
  \item \textit{Resolution}: actions (if any) taken to address the issue.
  \item \textit{Reason for resolution}: the reason that we identify based on the issue discussion for a certain resolution.
\end{itemize}

\subsection{Data Collection and Coding}
\label{sec:execution-plan}
After we identify the sample of subjects of our study according to the criteria presented in Section~\ref{sec:subjects}, we automatically collect the data for the variables \textit{Reporter, Labels, \#Comments, \#Discussants, Reporting date, Last active date}, and \textit{Status} using the appropriate GitHub APIs. The values for the other five variables \textit{Trigger, Privacy issue, Consent interaction, Resolution} and \textit{Reason for resolution} are obtained through manual coding. 

The actual manual coding was conducted using inductive coding \citep{Thomas.2006}. First, two of the authors independently coded the same 20\% of the collected data to create the code book. The two coders discussed their codes and ambiguities in the coding \rev{several times} in the first phase, \rev{which meant that a}fter around 8\% of the \rev{same} data was coded \rev{independently by the two coders}, a first harmonized code book was \rev{created. This code book was then} applied to code \rev{another 5\% of the same data independently by both coders, followed by a discussion on the codes and the codebook. After that the remaining 8\% of the same collected data were coded independently, using the improved code book}. This is common practice in other works in the area of usable security to ensure inter-rater reliability, used e.g. by \cite{Mayer.2021} and \cite{Pearman.2019}. It is, furthermore, part of the recommendations by \cite{Elder.1993wo}. After the same 20\% of the collected data was coded, Cohen's Kappa was calculated to measure inter-rater reliability (IRR) of the \rev{last increment, i.e. the last 8\% of the} coding. Since IRR was below $0.7$ after 20\% of data had been coded (\textit{Privacy issue}: 0.52; \textit{Consent interaction}: 0.35; \textit{Resolution}: 0.39), two additional 5\% increments of data were coded until an average IRR of 0.44 (\textit{Privacy issue}: 0.47; \textit{Consent interaction}: 0.28; \textit{Resolution}: 0.58) was reached. 

\begin{figure*}
\includegraphics[width=\linewidth]{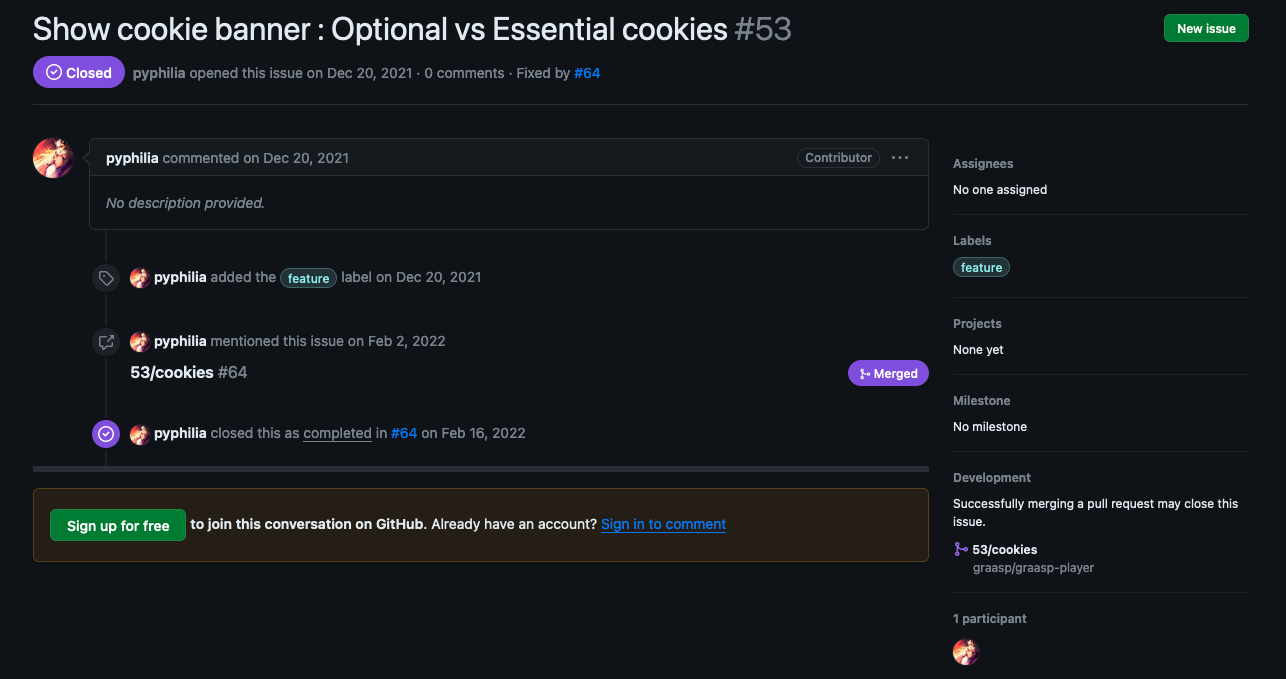}
\caption{Issue with only vague information provided.}
\label{fig:difficult-issue}
\end{figure*}

We found that the remaining differences were not due to ambiguities in the code book, but rather to ambiguities in the interpretation of the issues. For some issues, it was very difficult for outsiders to understand the comments and/or the proposed solutions. The descriptions were sometimes so vague that they could only be correctly interpreted by insiders (see Figure \ref{fig:difficult-issue} as an example for such an issue). Therefore, in the interest of time, we only refined the coding instructions and the remaining data was subsequently further coded by only one author. Additionally, both coders continued to discuss uncertainties and ambiguities in the coding. 

Furthermore, we determined at this stage that the addition of \textit{trigger} and \textit{reason for resolution} as additional variables is valuable and sufficiently detailed for the planned qualitative analysis. No further codes needed to be added at this stage and no further changes to the code book were made, indicating that the code book itself was comprehensive enough. \rev{To ensure the reliability of our codebook, 10\% of the material was additionally validated again by one of the authors who was not involved in the coding before. No notable differences were found (see Section~\ref{sec:reliability} for detailed information).} These methods were deemed sufficient given the exploratory nature of the study and the partial vagueness of the issues in our sample.

\subsection{Data Analysis}
\label{sec:analysis-plan}

In the following, we describe our qualitative and quantitative analysis. First, we describe the demographics of our population, followed by the quantitative analysis between the variables and the possibility to model predictive models for issue resolutions. Then\rev{,} we discuss the qualitative analysis we conduct based on what we observe during coding regarding the reasons for resolutions. 

\subsubsection{Descriptives of Demographic Data Pertaining to Research Questions}

We conduct an empirical analysis based on the data we collect. To understand what was reported, we report the results of the manual coding of the issue types, including the description and frequency of issue types related to personal data and data protection (\textbf{RQ 1}). In the same way, we provide data about the resolutions, i.e., the types of resolutions we observed including a description and the frequencies. We also present demographic data about our subjects to determine the prevalence of issues related to personal data and data protection (\textbf{RQ 2}). For this, we report the number of projects that match our criteria, the number of projects for which we identified at least one issue related to personal data and data protection, the number of overall issues of the projects, and the number of issues related to personal data and data protection identified by the keyword search. Moreover, we present for each keyword how many of the issues, which are identified using that keyword, had to be discarded during the sampling to measure the false positive rate of the keyword search. 

Then, we consider data for the individual perspectives of our research question \textbf{RQ 3}. To understand the reporters, we consider the number of reporters per different type of related issue. Similarly, we consider who joins discussions about issues related to personal data and data protection by reporting the numbers of discussants per different type. Furthermore, we report on the total number of issues related to personal data and data protection within each quarter in our study timeframe. As a small deviation from our protocol, we do not report data over time by reporter type or discussants of the issues, as the support within each quarter is to low, when we aggregate on such a fine-grained level.

\subsubsection{Relationships Between the Variables}
In addition to the individual reporting regarding our research questions, we also evaluate the relationships between our variables. We evaluate the relationship between the issue types and reporters through the absolute numbers of each reported issue type per reporter type. We augment this by an analysis of the cross-tabulation between the nominal reporter type and the numeric issues per type. We use the $\chi^2$ test with a significance level of $\alpha=0.05$ to get further information regarding the significance of the relationship. We restrict the statistical test to issue types which we observe at least 20 times, i.e., five times as often as the number of reporter roles. 

\subsubsection{Prediction of Issue Resolution}
Furthermore, we want to understand how the different aspects affect the resolution of issues related to personal data and data protection. For this, we try to predict the issue resolution as the dependent variable based on the other variables as independent variables, i.e., the \textit{Reporter, Discussants, Labels, \#Comments, \#Discussants, Trigger, Privacy issue}, and \textit{Consent interaction}. \rev{We note that while we use predictive models, we train them not with the target to actually predict outcomes later, but rather to understand correlations between variables. Due to this, we neither use train/test splits, nor do we need to address possible data leaks from the dependent variable. }

Since a discussion of an issue can have discussants of multiple roles \rev{(i.e. frequent reporter, one-time reporter, frequent committer, one-time committer)}, we encode this variable with four binary variables that mark for each role if there is a discussant with that role in binary form (i.e., 1 if the discussant role is present, 0 if not). Consequently, we have eleven independent variables for these models: two numeric variables (\textit{\#Comments} and \textit{\#Discussants}) and nine nominal variables (\textit{Reporter, each of the four Discussant types, Labels, Trigger, Privacy issue}, and \textit{Consent interaction}). \rev{We use one-hot encodings for the non-binary nominal variables, i.e., for \textit{Reporter, Labels, Trigger, Privacy issue}, and \textit{Consent interaction}.} We take the pattern from \cite{Tunkel2022} and create multiple models to understand the relationships between our variables: 1) a multinomial logit model to understand the (linear) relationship between the independent variables and the odds of the resolution; 2) a decision tree to understand if we can find a description based on Boolean rules for the resolution; and 3) a random forest to understand if a powerful non-linear approach can model the relationship. This multi-perspective approach means that we combine less powerful models that are easy to interpret (linear model for coefficient relationships, rule-based models to understand how concrete values behave) with a more powerful non-linear model to avoid assuming the lack of a relationship as a consequence of underpowered modeling techniques. Based on \cite{Bujang2018}, we estimate that we require $n=100+50\cdot \#\text{independent variables}=100+50 \cdot 11 = 650$ issues related to personal data and data protection for the multinomial logit model. In the absence of similar rules for the other models, we use 650 as required sample size for our study. 

\paragraph{Prediction Based on Multinomial Logit Model} For each independent variable, we compute the average marginal effects, i.e., the average slope of the logistic function when the variable values change. These slopes can then be interpreted as the change in the probability of the resolution when the independent variable value changes. We determine which marginal effects are significant using a significance level of $\alpha=0.05$ using the $z$-values computed based on the standard error of the estimated marginal effects as test statistics. Furthermore, we report McFadden's adjusted $R^2$~\citep{McFadden1974} to report the goodness of fit of the model. This helps us to further understand the reliability of the odds, as the coefficients of a model with a poor fit are less reliable.

\paragraph{Prediction Based on Decision Tree} The decision tree can directly work with the nominal data and does not require one-hot encoding. We use a CART decision tree~\cite{Breiman1984} with Gini impurity as splitting criterion. The choice of splitting criterion has been shown to not have a large impact on the resulting trees \citep[see, e.g.,][]{Raileanu2004}. We do not restrict the tree depth and conduct a manual analysis of the resulting decision tree. Thus, instead of using the overall accuracy to determine the quality of the model which may have problems with overfitting, we rather consider the individual data partitions at the nodes of the decision trees, as this allows us better and more fine-grained insights. We consider which decisions were made, how the decisions help to decide for specific resolutions, as well as the general support of the decisions, i.e., the amount of data used for the decision and within the resulting subsets.

\paragraph{Prediction Based on Random forests} Random forests~\citep{Breiman2001} are consistently among the best performing machine learning models for smaller tabular data sets~\citep{FernandezDelgado2014}. A random forest determines a non-linear relationship between the dependent variable and the independent variable through an ensemble of decision trees, where each decision tree is trained on a subset of the data and variables. In contrast to the decision tree, we cannot feasibly manually analyze a decision tree to understand the relationship, as we would have to consider hundreds of trees. Instead, we use this analysis to augment our insights from the less powerful but interpretable decision trees. Concretely, we calculate the feature importance, which measures how much each feature contributed to the reduction of the Gini impurity that is observed at the leaf nodes of the trees, averaged over all trees. 

We augment the above analyses for the relation of the resolution with an analysis of the confusion matrices, to understand if the models are better at modeling some resolutions than others. Furthermore, we compute the correlations with Spearman's $\rho$ between all variables, as this allows us to understand interactions between variables within the models. 

\subsubsection{Reasons for Choosing Specific Implementations \& Solutions}

The qualitative analysis aims to identify the reasoning for choosing specific implementations for specific issues related to personal data and data protection, and to which resolutions these implementations lead (\textbf{RQ 4}). Additionally, the coders consider any further aspects of note that come up as part of this exploratory analysis. Count data of all codes is reported (as opposed to percentages) to avoid over-generalizing. The individual concepts uncovered in the coding will be illustrated using quotes from the data.

\subsection{Deviations from pre-registered study protocol}
\label{sec:deviations}

Due to our better understanding of the underlying data, we deviated from the pre-registered study protocol in several small aspects, which we summarize below. 
\begin{itemize}
    \item We added the independent variables \textit{trigger} and \textit{reason for resolution}. \rev{Within the qualitative coding, we} quickly discovered that they often saw relevant information regarding why issues were reported, but that they had no place within the current variables to encode this information. To address this issue and enrich our analysis, we added these variables.
    \item We dropped the independent variable \textit{status}. Since many of the resolutions directly consider if an issue was addressed and, hence, closed, this would have been an information leak for our analysis of predictions of the issue resolution. Removing this variable addressed this issue without losing other valuable information. We note that the labels may also leak information about the issue resolution, especially when they are about the progress. However, the labels encode a lot more information and are also a less reliable information leak, which is why we decided to keep the labels for our analysis as planned. 
    \item We dropped the independent variable \textit{last activity}. This had no impact on our analysis. We initially planned to use this variable for the analysis over time as well, but this turned out to be redundant to the analysis over time based on the reporting date. 
    \item We dropped the reporting of the evolution of the reporter roles and discussants over time further aggregated by issue types. The support in the data (i.e., non-zero values) when aggregating on this level was not sufficient for almost all issue types, rendering such an analysis useless, without at least 5 times as much data.
    \item We analyze the multinomial logistic regression based on marginal effects and not based on the coefficients. The reason is that the significance of coefficients of a multinomial logit model is computed against a reference level. However, we do not have a reasonable reference level which we could have used within our encoding of the resolution types. Using marginal effects, we can conduct a similar analysis without this problem.
\end{itemize}

\section{Results}
\label{results}

In this section, we report the demographics of the data privacy issues on GitHub, how the different developer roles are related to the types of privacy issues, how our variables are correlated, as well as how the variables are related to the resolution of an issue by considering predictive models. 

\begin{figure}
\centering
\includegraphics[width=\textwidth]{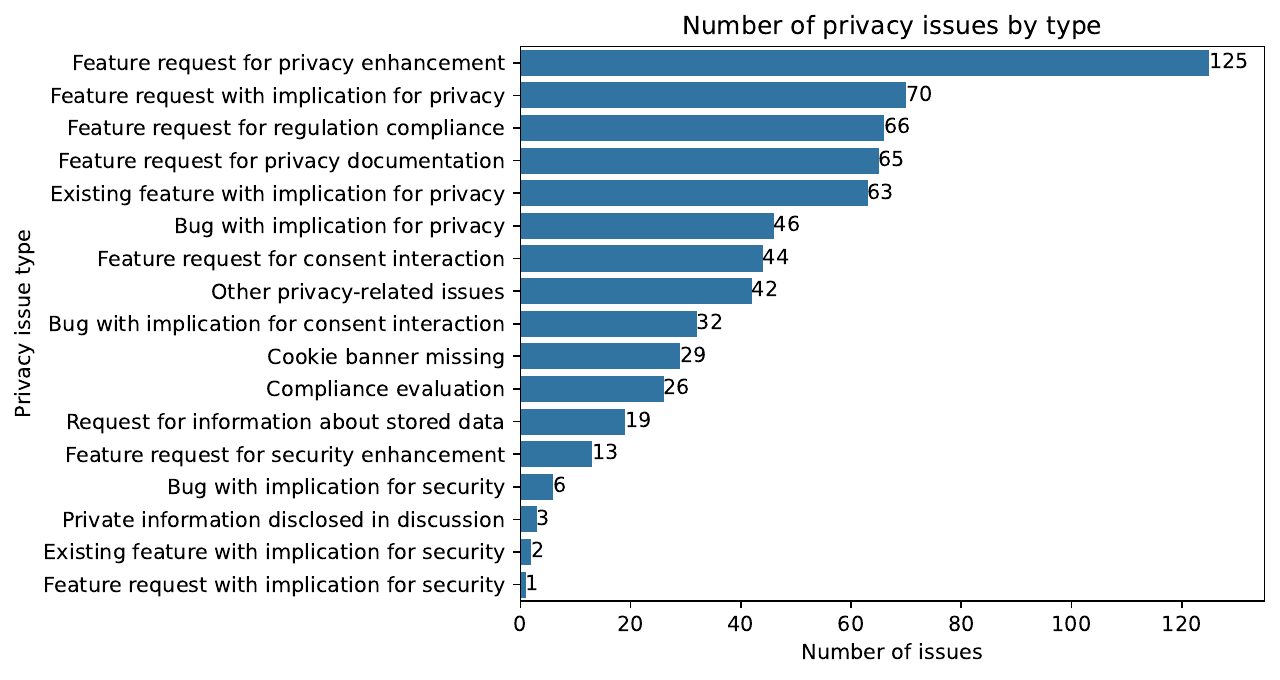}
\caption{Overview of privacy issues and the counts of how often they were observed.}
\label{fig:privacy-issue-counts}
\end{figure}

\begin{figure}
\centering
\includegraphics[width=\textwidth]{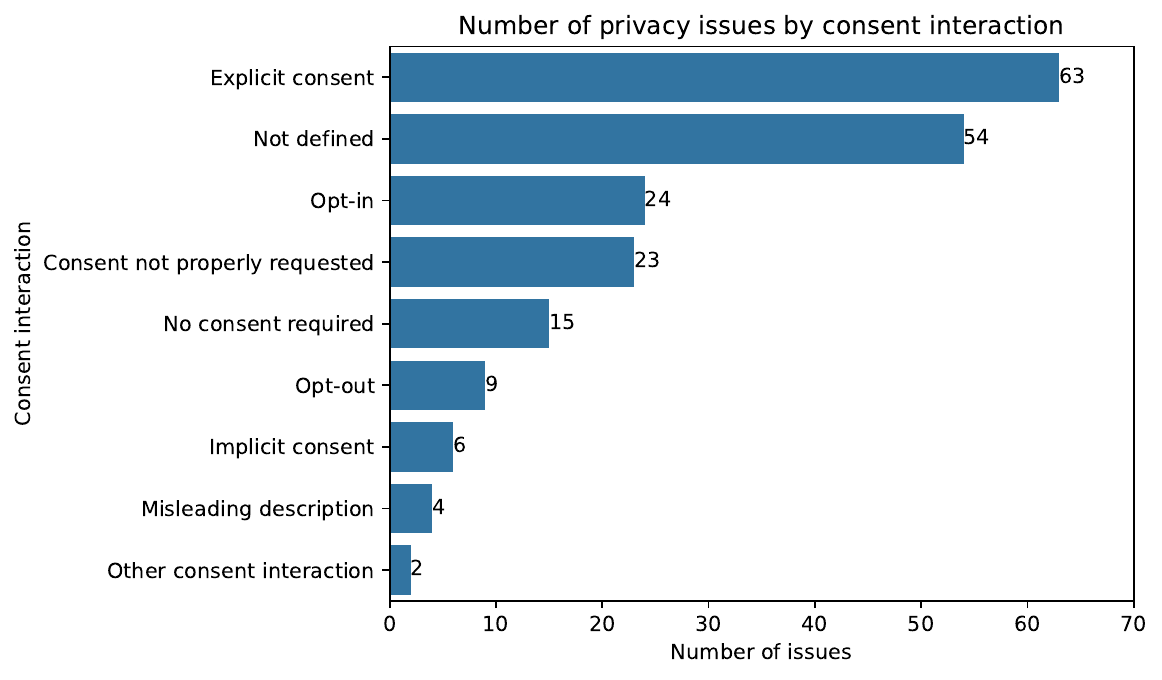}
\caption{Overview of consent interactions discussed within privacy issues. Omits 452 issues for which the consent interaction was not relevant.}
\label{fig:consent-interactions}
\end{figure}

\begin{figure}
\centering
\includegraphics[width=\textwidth]{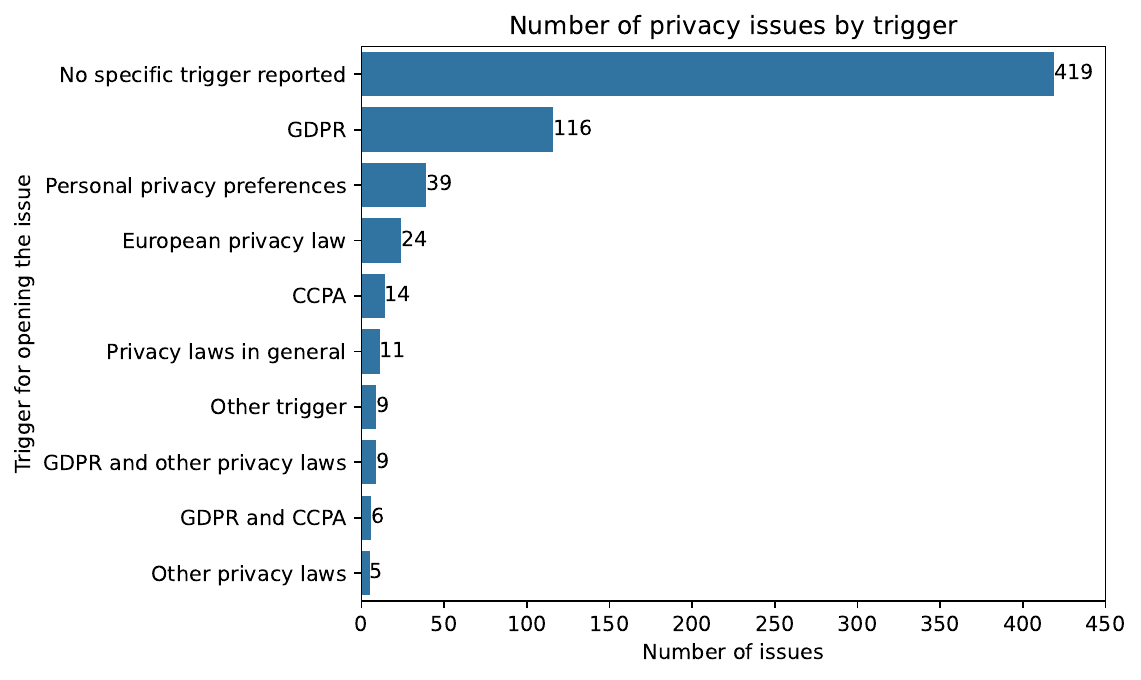}
\caption{Overview of trigger events for the creation of privacy issues.}
\label{fig:trigger-issue-counts}
\end{figure}

\begin{figure}
\centering
\includegraphics[width=\textwidth]{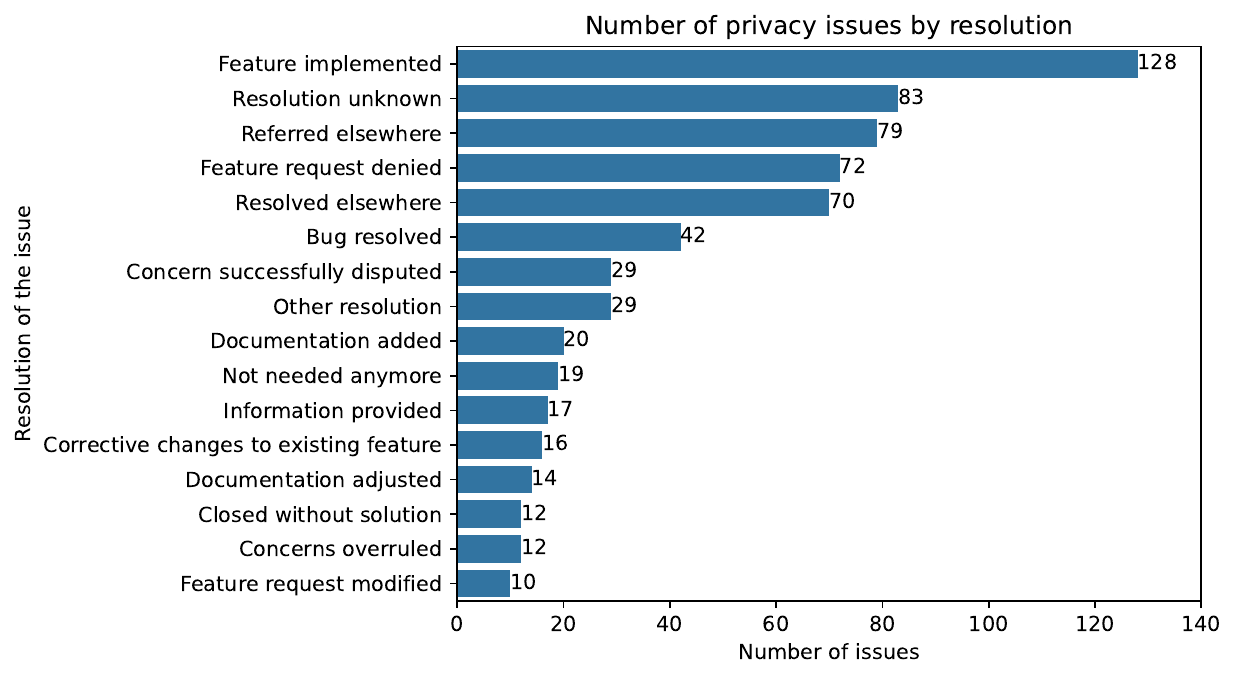}
\caption{Overview of resolutions of privacy issues.}
\label{fig:resolution-issue-counts}
\end{figure}

\begin{figure}
\centering
\includegraphics[width=\textwidth]{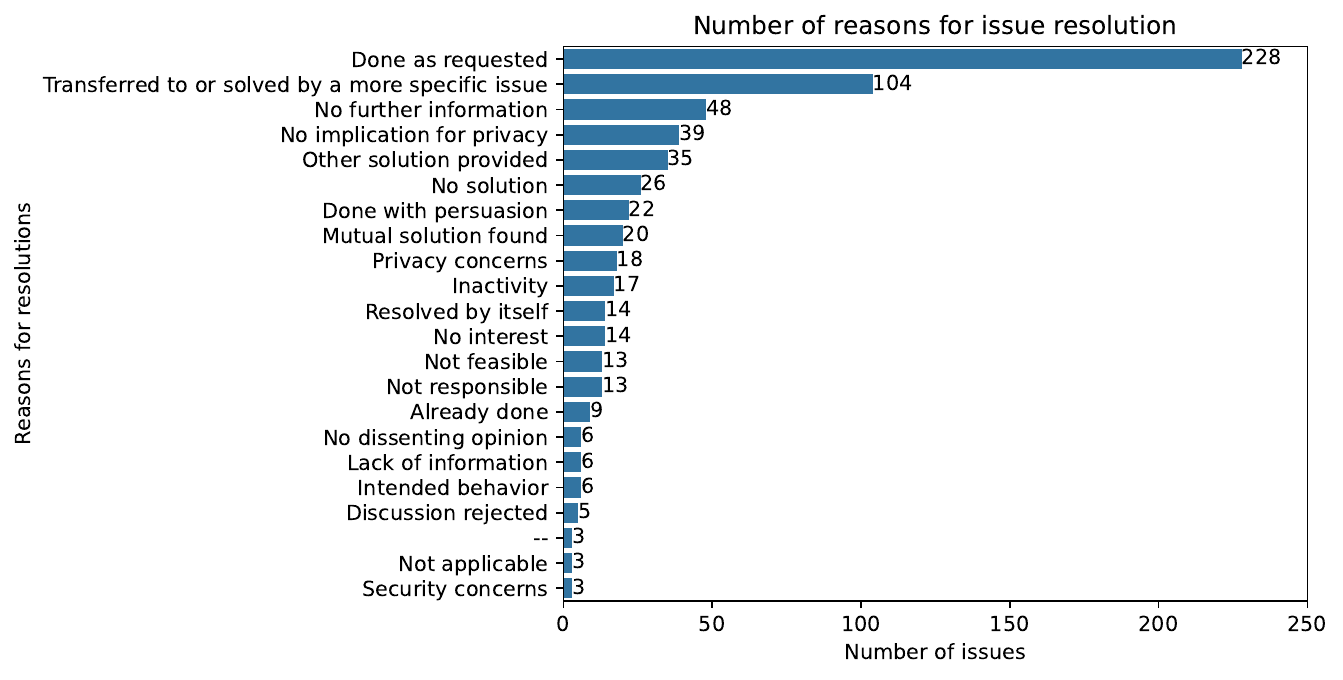}
\caption{Overview of reasons for issue resolutions. There are three issues, which were deleted from GitHub when we checked for the reason marked as ``--''.}
\label{fig:resolution-reason-issue-counts}
\end{figure}

\begin{figure}
\centering
\includegraphics[width=\textwidth]{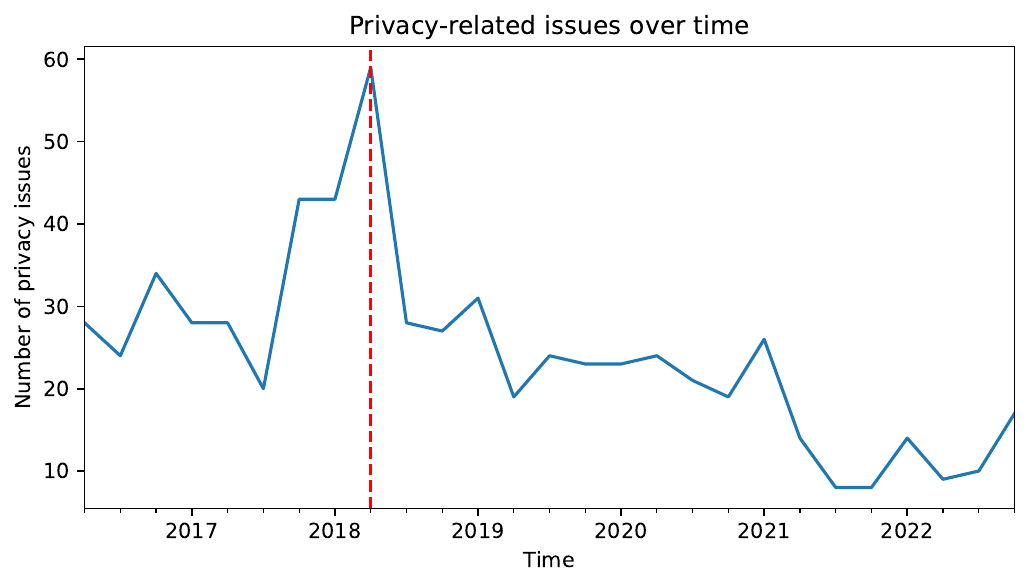}
\caption{Number of privacy issues over time. The dashed line is in the \rev{first} quarter of 2018 where both the GDPR and CCPA were first enforceable.}
\label{fig:issues-over-time}
\end{figure}

\subsection{Demographic data}
\label{sec:results-demographics}

Overall, we found 12,606 issues from 5,069 projects that met our inclusion criteria defined in Section~\ref{sec:subjects} and matched at least one of the keywords. We observed an overall false positive rate of our keyword search of 63\% when we manually validated whether the issues were indeed about data protection. The main driver for this high false positive rate were terms that are also used outside the context of data protection, like ``data sharing'' and ``fingerprinting'', but also some more specific terms like ``cookie prompt'' and ``cookie notice''. The reason for the latter is that cookies are also often discussed independent of their data protection implications, e.g., when it comes to implementing the logic of a website. Assuming that the false positive rate is the same for all 12,606 issues, we expect that there are $12,606\cdot 0.37 = 4,664$ issues about data protection within the 5,069 projects.

\rev{\paragraph{Variable Reporting Date}
As a first demographic factor, we considered trends over time. Figure~\ref{fig:issues-over-time} shows the overall number of privacy issues in our sample through the years. Except for a peak at the time when GDPR and CCPA were first enforceable in the first quarter of 2018, the reporting activity of privacy-related issues is fairly constant until the end of 2020. Afterwards, the reporting is a bit less, which indicates that the discussions regarding data protection in our included projects are getting less in the last two years. We also analyzed the trend over time for the issue types but did not observe any notable patterns.}

\paragraph{\rev{Variable Privacy Issue}} 
Figure~\ref{fig:privacy-issue-counts} shows the counts of the privacy issues we found for each type\rev{\footnote{See Appendix \ref{sec:codebook} for an explanation of the different issues types.}}. The data shows that the issues regarding data protection are considered from different angles and affect all parts of the software development life cycle: the addition of new features specific to data protection (n=125), the discussion of data protection implications for both new (n=70) and existing features (n=63), the documentation of data protection aspects (n=65), as well as the correction of existing features to fix data protection issues (n=46) or to be compliant with regulation (n=66). A notable aspect that we observed was that many of these issues are specific to user interactions for containing consent (n=32), incl. missing cookie banners (n=29). Moreover, the evaluation of compliance is also a topic discussed within issues (n=26). Data security aspects are also reported through feature requests specific to data security (n=13), the discussion of implications for new (n=1) and existing features (n=2), as well as data security bugs (n=6). We also found few issues (n=3) in which private information was revealed as part of the issue, which was subsequently discussed directly within the issue. Moreover, there were even requests for information about stored data from individuals (n=19) directly on GitHub. Beyond this already very diverse reporting of issues related to data protection, we also found 42 issues that are otherwise related to data privacy, e.g., bugs caused by privacy settings or features, or privacy features that limit functionality (e.g. a cookie banner disturbing UX). Among them were also four issues that stood out from any category: One was a bug report, where the reporter did not want to publicly post a debug output on GitHub due to privacy concerns.\footnote{\url{https://github.com/hashicorp/terraform/issues/11331}} The second and third one were probably bulk messages sent to several projects demanding to leave GitHub\footnote{\url{https://github.com/netblue30/firejail/issues/2895}} or offering GDPR compliance support.\footnote{\url{https://github.com/arx-deidentifier/arx/issues/126}} The fourth issue was a ``Do not trust this site'' - warning by a bot of the United States Digital Service (USDS).\footnote{\url{https://github.com/department-of-veterans-affairs/vets-website/issues/2110}}

\paragraph{\rev{Variable Consent Interaction}}
\rev{Figure~\ref{fig:consent-interactions} depicts the consent interactions that were discussed as part of data protection issues\rev{\footnote{See Appendix \ref{sec:codebook} for an explanation of the different codes for consent interaction.}}. Consent was not discussed for 452 issues in our data set\rev{, mainly because the issue was not related to a feature where user consent was needed}. Furthermore, there were 54 issues, in which consent was discussed broadly, but without any \rev{specifications} about \rev{concrete interactions} to obtain consent. When consent is discussed, \rev{e.g., in the context of cookie banners or when an application is installed for the first time,} this is mostly about the type of consent. On the one hand, whether this should be explicit (n=63), opt-in (n=24) or opt-out (n=9), or, on the other hand, whether this should be implicit (n=6), meaning just a not\rev{e} without any options to set preferences. We note, that explicit consent is a lot more often \rev{discussed} than implicit \rev{consent}. There are also cases where it was decided that no consent is required (n=15), e.g., due to legitimate interest \rev{or because no user data are stored}. Other times, issues are about improper consent request\rev{s} (n=23) or the description of a consent interaction \rev{that} was misleading or inaccurate (n=4), e.g., when an app asked for the permission to use GPS data ``once the location changes significantly'', but it was not clearly defined whether ``significant change'' means a few hundred meters or several kilometers.\footnote{\url{https://github.com/nextcloud/ios/issues/324}} Finally, there are two cases in which other aspects of consent interactions were discussed, i.e., the removal of cookie banners\rev{. More specific, o}ne issue where the cookie banner should be removed without giving further information about how the cookie banner had looked before or why it should be removed\footnote{\url{https://github.com/chanzuckerberg/cellxgene/issues/2235}}. And the second where analytics should be implemented but not in the EU so that no cookie banner would be necessary.\footnote{\url{https://github.com/StampyAI/stampy-ui/issues/81}}}

\paragraph{\rev{Variable Trigger}}
Figure~\ref{fig:trigger-issue-counts} shows the counts of the triggers that led to the reporting of an issue\rev{\footnote{See Appendix \ref{sec:codebook} for an explanation of the different codes for trigger.}}. For most of the privacy issues, the reporters did not specify a concrete reason that triggered the reporting of the issues (n=419). Of the reported reasons, the GDPR seems to be the clear driving factor (n=116), especially when we consider that sometimes the reason is just given as European privacy law in general without specifically mentioning the GDPR (n=24), and that the GDPR is also sometimes the reason together with other privacy laws (n=9) or the CCPA (n=6). The second largest driving factor of reporting of issues are personal preferences of the reporters (n=39). The CCPA (n=14), privacy laws in general (n=11), and other privacy related laws (n=5) are also mentioned as reasons, though not that often. Additionally, we also found other reasons for reporting (n=9) that were not purely data protection regulations, e.g., FERPA (Family Educational Rights and Privacy Act) or COPPA (Children's Online Privacy Protection Act), requirements from other services, like the Google Analytics Terms of Service (ToS) or the Apple App Store, or in general \rev{``}data protection\rev{''} or \rev{``}the right to be forgotten\rev{''}.

\summaryboxnew{\rev{In total, we can see that topics related to personal data and data protection are discussed in various contexts throughout the software development process, with news features to enhance users' privacy having the greatest share. Since only a few issues address user consent, consent interaction was not discussed for the majority of issues.  We also observe that for the majority of issues no dedicated trigger could be detected. If an issue contained a specific trigger, this was mostly GDPR. The fact that data protection regulations are a driver for issues related to personal data and data protection is also reflected in the trends over time, where we could see an increase in issues from the second quarter of 2017 that peaks in the first quarter of 2018.}}

\paragraph{\rev{Variable Resolution}}
Figure~\ref{fig:resolution-issue-counts} gives an overview of the resolutions to the data protection issues\rev{\footnote{See Appendix \ref{sec:codebook} for an explanation of the different codes for resolution.}}. The most common resolution of issue types is that they are addressed through implementing features (n=128), because they are resolved elsewhere (n=79), the underlying bug is fixed (n=42) or other corrections of features that the developers did not consider to be bugs (n=16). We also observed that a feature request was modified to be compliant with e.g. regulations (n=10), that documentation\rev{, e.g., privacy documentation or consent interaction,} was added (n=20), adjusted (n=14), or the desired information regarding privacy-related aspects was provided (n=17). There were also cases, in which the original concern from the reported issue was successfully disputed (n=29) \rev{by mutual agreement and consequently no changes were made}, as well as cases where the concerns were acknowledged but overruled without addressing them (n=12). Moreover, in many cases the developers referred the request elsewhere (n=79), e.g., another bug tracker tool or another, more appropriate issue, or closed the issue without being able to provide a solution (n=12). In some cases, the data protection issue became irrelevant due to other changes (n=19), e.g., if a feature is already designed in a way that does not infringe data protection regulations\footnote{e.g., \url{https://github.com/decidim/decidim/issues/2266}}, or if a feature was dismissed or (already) changed in a newer version so solving the bugs or implementing privacy enhancing techniques was not relevant anymore.\footnote{e.g. \url{https://github.com/inaturalist/inaturalist/issues/1012}} Additionally, there were several other resolutions (n=29), e.g., if the code was not applicable \rev{when the privacy concern was about} personal data \rev{being} revealed in a discussion. Or when the issue was solely for \rev{internal} documentation \rev{or notes}, if information were not provided as requested or discussions were denied, or if another solution or workaround was provided without any changes to the application. Finally, we failed to determine the resolution for 83 issues, e.g. when the issue was closed without a recorded solution other than the automatic message \rev{that the issue is closed}. 

\paragraph{\rev{Variable Reasons for Resolution}}
Figure~\ref{fig:resolution-reason-issue-counts} depicts the reasons for issue resolution, which give us further insights into the \rev{motivations for resolving an issue}\rev{\footnote{See Appendix \ref{sec:codebook} for a the explanation of the different codes for the variable reasons for resolution.}}. In most cases, issues were resolved as requested by the reporters without further discussion (n=228), resolved themselves (n=14), or were already done (n=9). In other cases, this required persuasion, i.e. discussion, (n=22), a different solution (n=35), sometimes achieved by mutual agreement in discussion (n=20), or when there was no further discussion and, thus, no disagreement of opinions regarding a provided solution or explanation (n=6). Reasons for not resolving issues were that developers saw no implications for privacy (n=39), that they could not offer a solution (n=26), that a solution was not feasible (n=13), that developers showed no interest (n=14), or the issue went stale due to lack of information (n=6). There were security concerns (n=3), but also sometimes disagreements because the behavior was intended (n=6). In some cases, the discussion was discontinued by the developers and limited to collaborators only (n=5). Other issues were going stale due to inactivity (n=17). Moreover, in line with the resolution of the issues, there is also a large part of issues that are transferred or solved elsewhere (n=104).

\summaryboxnew{\rev{In total, we can see that in most of the cases developers addressed the issue either by implementing a feature, resolving it in another issue, or by correction, e.g., fixing the bug that caused the issue. We could also observe that in many instances, no resolution is provided, or the issue is referred elsewhere and no dedicated resolution for the issue at hand is provided. Implementation is mainly done as requested, but for a lot of issues we could not identify specific reasons, because the issue was missing information about this. If a feature request or bug fix was denied, this was mainly because the issue resolved itself, was already done, or developers saw no implication for privacy.}}

\subsection{Relationship between developer roles and issue types}
\label{sec:results-relationships}

\begin{figure}
\centering
\includegraphics[width=\textwidth]{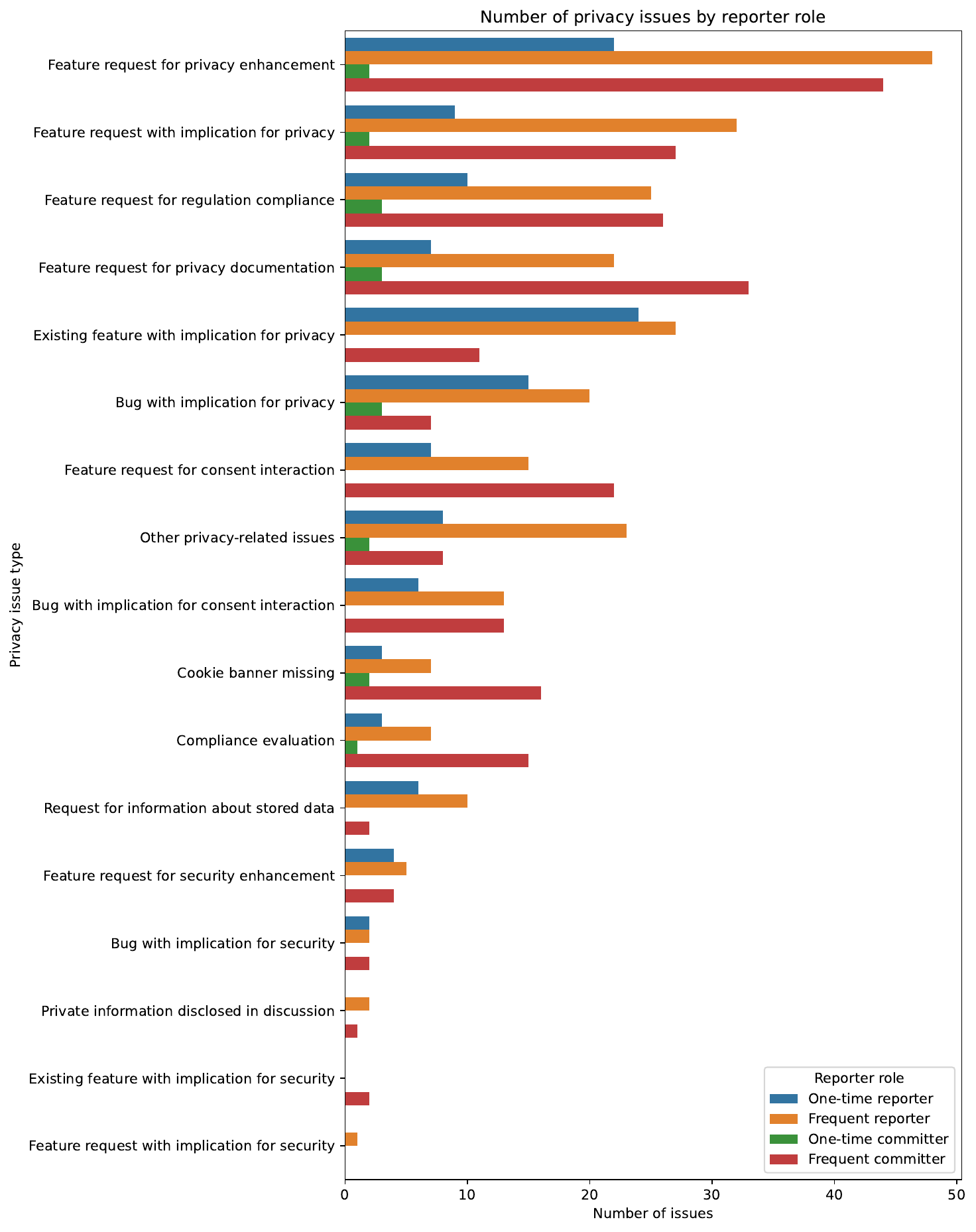}
\caption{Overview of the reporter roles per privacy issues type.}
\label{fig:reporter-issue-counts}
\end{figure}

\begin{figure}
\centering
\includegraphics[width=\textwidth]{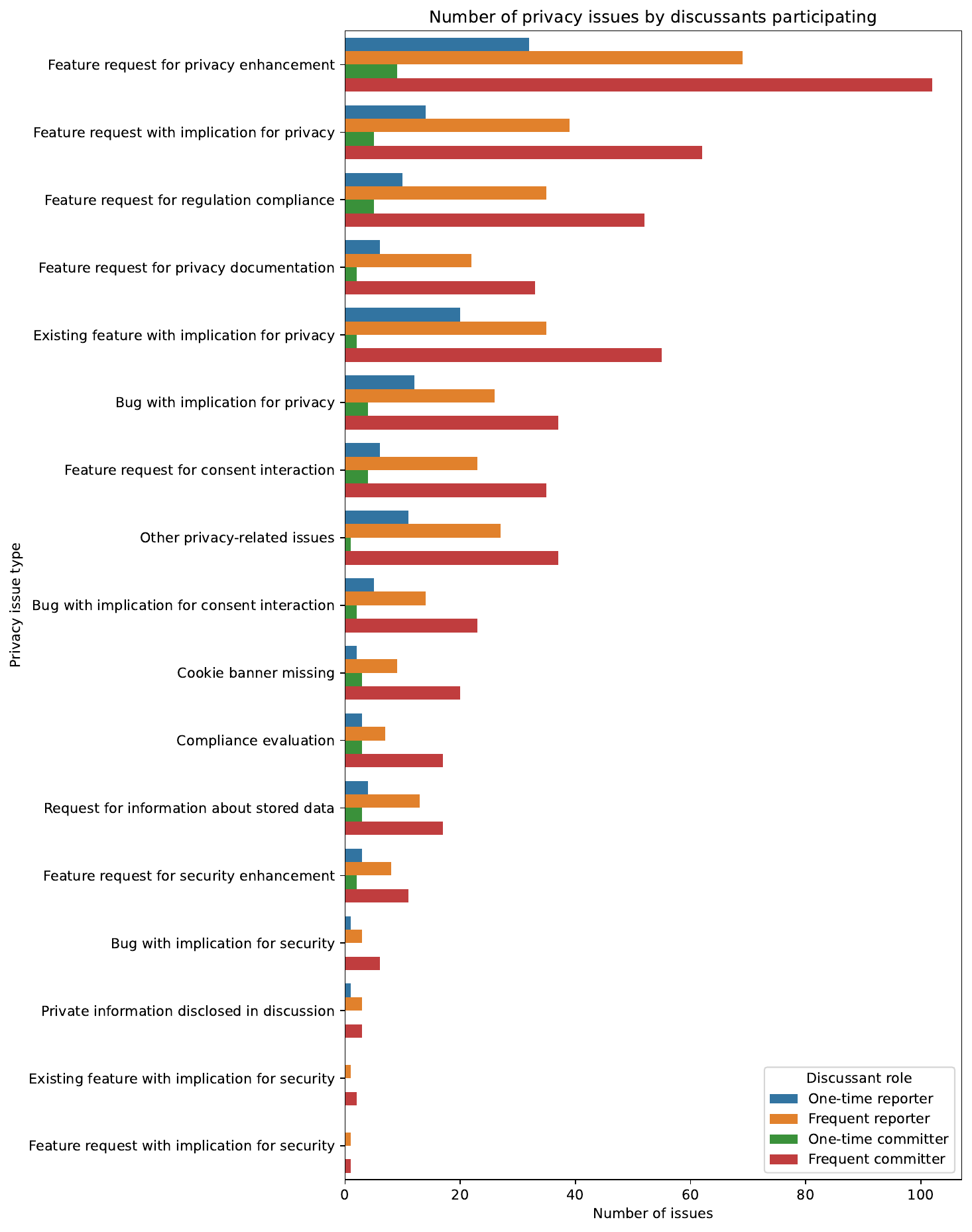}
\caption{Overview of the discussant roles per privacy issue type.}
\label{fig:discussant-issue-counts}
\end{figure}

Figure~\ref{fig:reporter-issue-counts} depicts the roles of the issue reporters. This data is reported on a subset of 636 issues, because the role of the reporter is unknown for the other issues since their accounts were deleted from GitHub at the time of our data collection. For all issue types, the frequent reporters and frequent committers\rev{, i.e., persons who are frequently active in the issue and probably part of the development team,} dominate \rev{by} report\rev{ing} the most issues. However, there are also many one-time reporters, \rev{i.e. reporters who are only active within a single issue,} especially for issues regarding the improvement of privacy through new features, updates of existing features, or bugs. Moreover, we observe that issues regarding how consent is implemented -- including cookie banners -- come mostly from frequent \rev{committers}, indicating that this is commonly considered to be rather a development issue. The role we observe most seldom, are one-time committers\rev{, i.e. persons who only contribute to the code within a single issue}. They appear fairly randomly over the issue types, indicating that sometimes people with the necessary skill\rev{s} invest the effort to directly address a data protection issue in an otherwise unfamiliar project, but these are rather outliers. When we conduct a chi-squared test based on the cross-tabulation between \rev{the reporter roles and} all privacy issue types that appear at least 20 times, we find that the relationship between these two variables is likely significant (p$<$0.001). This indicates that the above-discussed differences in the reporting behavior are not just due to the absolute number of reporters, but rather a real effect due to the differences in the involvement in the projects. 

Figure~\ref{fig:discussant-issue-counts} \rev{depicts the roles of the discussants of an issue. This} paints a \rev{slightly} different picture\rev{:} Here, the frequent committers are most active for all issue types, followed by frequent reporters. One-time reporters \textit{and} one-time committers play only a minor role. Thus, while people frequently \textit{report} only a single issue, they do not often join a discussion. When we look at the data in depth, we also observe that for most issue types the number of frequent committers is about 1.5 times higher than the number of frequent reporters. However, there are also some issue types where there are, relatively, fewer frequent reporters involved, i.e., less than half as often as frequent committers (e.g., for missing cookie banners and compliance evaluations). 
This is in line with the reporting of such issues, where the frequent committers are also more involved. When we conduct a chi-squared test based on a cross-tabulation between the privacy issue types and the different discussant roles involved, we again find that this relationship is likely significant (p=0.005). Same as above, this indicates that the differences in discussion activity are \rev{related} to the difference in the involvement in the projects by the different roles.

\summaryboxnew{\rev{In total, we can summarize that frequent committers and frequent reporters, i.e. people who are frequently active within the project, dominate both the reporting and the discussion of issues. But while one-time reporters are active in reporting issues, especially issues asking for features to enhance users' privacy, as well as reporting bugs or existing features that have an impact on user's privacy, they seldom contribute to discussions. One-time committers are least active in both reporting and discussing issues.}}

\subsection{Correlations between variables}

\begin{figure}
\centering
\includegraphics[width=\textwidth]{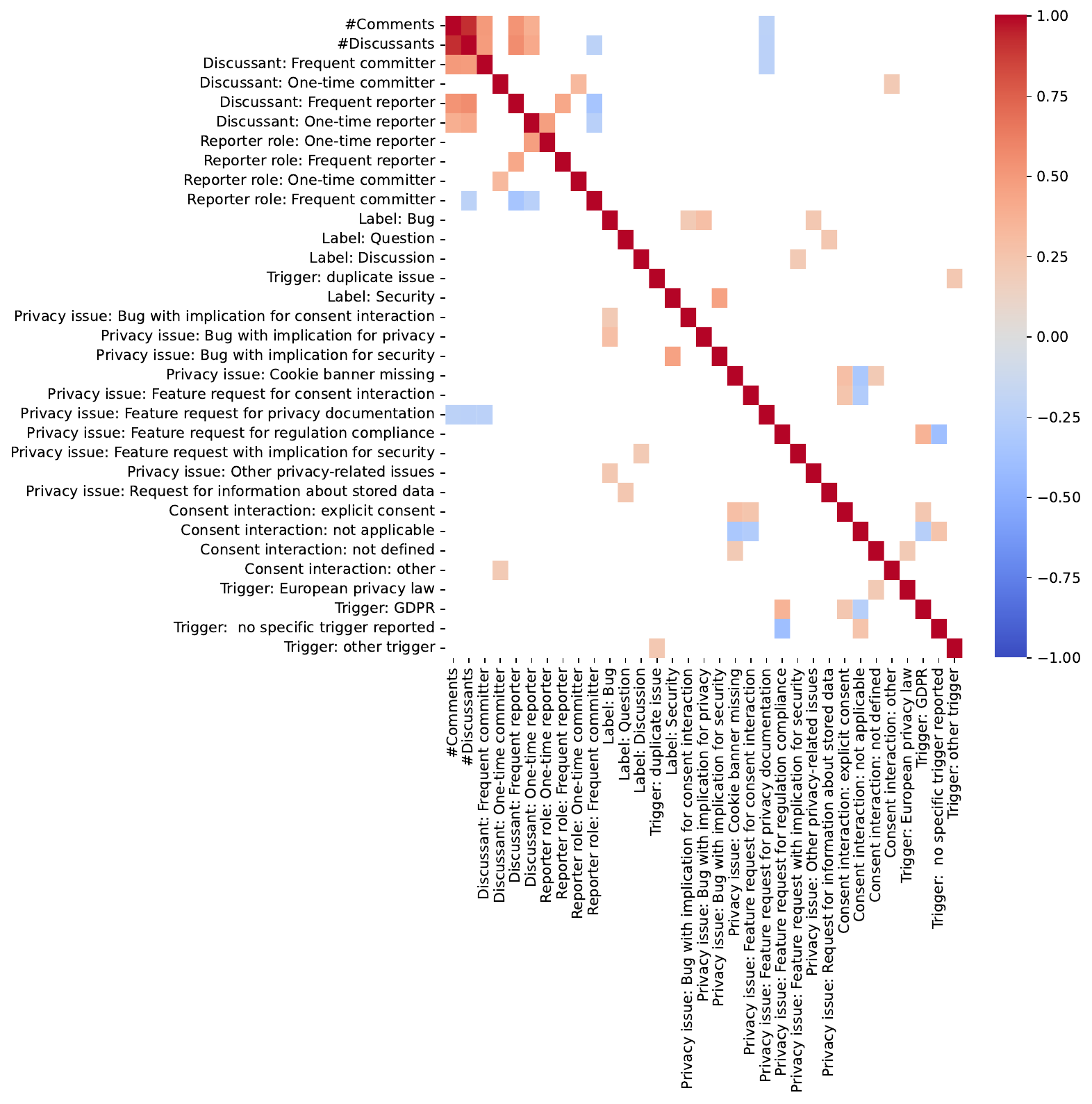}
\caption{Heatmap of Spearman's $\rho$ correlation between variables. Correlations less with absolute values are not depicted. Variables that do not have any absolute correlations above 0.2 are dropped. Correlations between the one-hot encoded features of the same variable are set to zero.}
\label{fig:correlations}
\end{figure}

Figure~\ref{fig:correlations} shows the correlations between our independent variables. As can be expected, the number of comments in an issue have a positive correlation with the number of discussants ($\rho$=0.92). Both variables also have positive correlations with the involvement of frequent committers, frequent reporters, and one-time reporters ($\rho \in [0.38, 0.56]$) in the discussion. We note that there is no such correlation to the one-time committers and the reporter roles. However, with frequent committers as reporters, we actually observe a negative correlation with the number of discussants ($\rho=0.23$). \rev{Furthermore, we see a }negative correlation between frequent developers as reporters and non-committers joining the discussion ($\rho \in [-0.3, -0.24]$). For the other roles, we rather observe moderate positive correlations between developers in the roles one-time committer, frequent reporter, and one-time reporter and their involvement in the discussions and reporting of issues ($\rho \in [0.32, 0.48]$). Another aspect is that there seem to be fewer discussions with less people involved for issues regarding the privacy documentation, which indicates that such issues either may not require a lot of discussion or are deemed less interesting. 

Moreover, consent interactions about explicit consent have a positive correlation with GDPR as a trigger ($\rho=0.23)$ as well as the privacy issue types missing cookie banner ($\rho=0.28)$ and feature requests for consent interactions ($\rho=0.24$). Vice versa, there is a negative correlation between consent not being applicable and these aspects ($\rho \in [-0.32, -0.26]$). Furthermore, regulation compliance issues are positively correlated with GDPR as a trigger ($\rho=0.36)$ and negatively correlated with no trigger ($\rho=-0.39$). Beyond this, we observe some weak correlations, without notable patterns, e.g., between labels and issue types.

\summaryboxnew{\rev{By looking at the correlations between our independent variables we see that only few variables are highly correlated. Most notably, we see positive correlations between the number of comments and the number of discussants, which makes sense since the people are involved in a discussion, the more comments can be expected. We also observe positive correlations between the discussant roles, and the number of comments and discussants, meaning that issues have more comments and discussants, when either frequent reporter, frequent committer or one-time reporter are part of the discussion. However, this is not true when one-time committer are involved in the discussion. We could also see that, when frequent
committers report data protection issues, there are fewer people involved in the
discussion, i.e., it has a negative correlation to the number of comments. This could be an indication that such issues are rather handled by
the reporting developers and/or other members from the project’s core team,
i.e., a small group of people. Furthermore, we see positive correlations between issues that deal with missing cookie banners or other types of consent interaction, and GDPR as trigger, which provides a strong indication that GDPR is a key driver \rev{for issues about} compliance with data protection regulations.}}

\subsection{\rev{Relationship between resolutions and resolution types}}

\begin{figure}
\centering
\includegraphics[width=\textwidth]{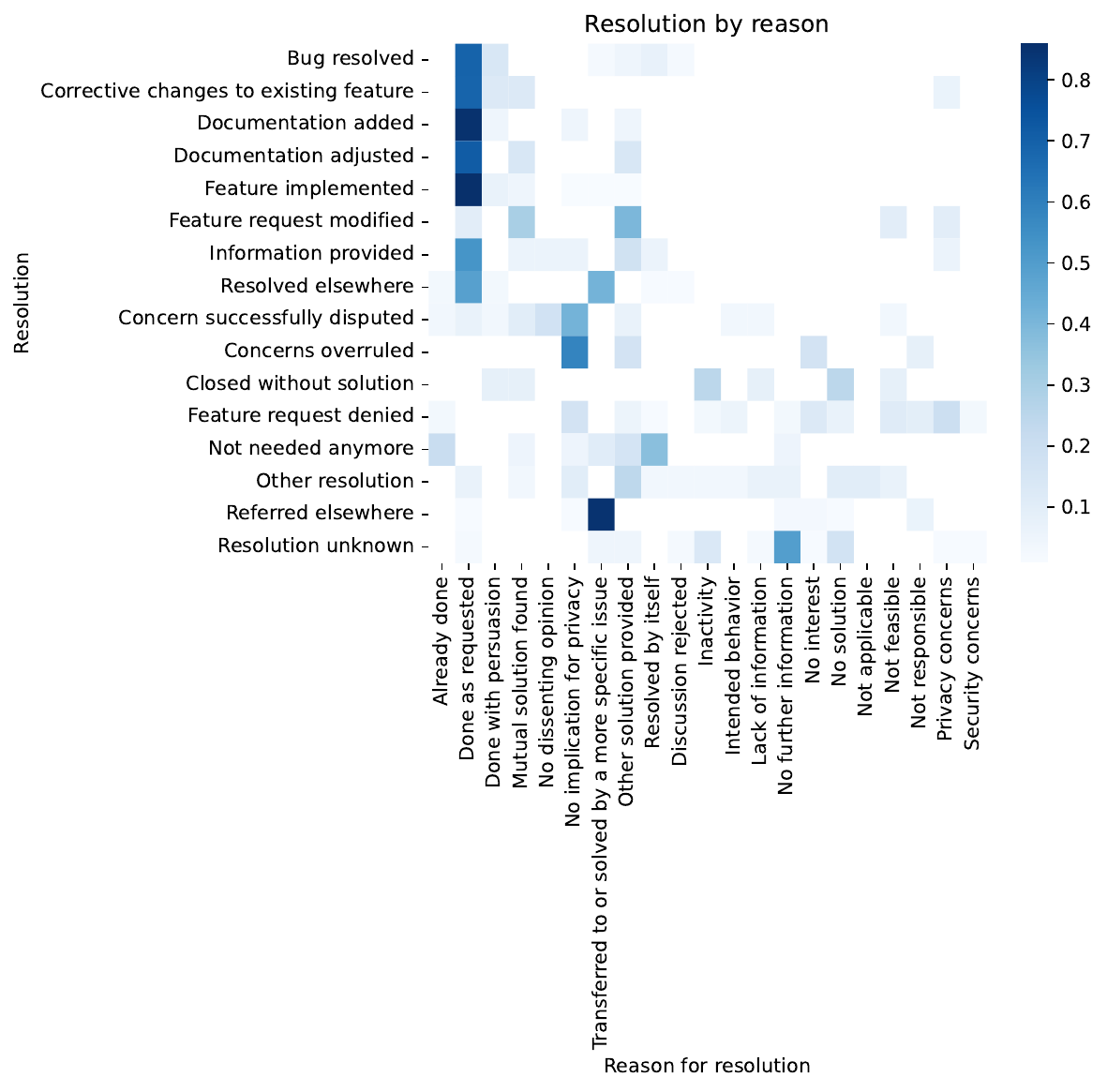}
\caption{\rev{Heatmap of the relationship between resolution types and resolutions.}}
\label{fig:resolutions_vs_types}
\end{figure}

\rev{Figure~\ref{fig:resolutions_vs_types} shows how the reasons for the resolutions we determined relate to the types of the resolutions. As can be seen, if issues were resolved, this was mostly done as requested, even though this sometimes required persuasion or finding a different solution then suggested. In case issues disputed or overruled, this was usually because the developers though no privacy concerns, though the developers sometimes also considered alternative solutions instead. For issues that were closed without solutions are rejected, the reasons vary strongly between all reasons for rejections we identified, i.e., there is no common reason for rejection.}

\summaryboxnew{\rev{When privacy issues are resolved, this is usually done as requested, issues are overruled based on a lack of actual privacy concerns. The reasons for not addressing valid issues vary broadly, e.g., lack of interest, required information, or feasible solution.}}

\subsection{Results for resolution prediction}

When we model the issue resolution with a multinomial logit model, we unfortunately have to drop all variables which are non-zero in less than 25 of the issues, i.e., about 4\% of the overall data. This is required because of the many one-hot encoded categorical features, in which the support for some categories is not large enough, leading to a badly conditioned problem for which the solution does not convert. These variables are listed in Appendix~\ref{sec:dropped-variables}. 

\rev{Figure~\ref{fig:confusion-matrix-logit} shows the confusion matrix of the multinominal logit model.} The model achieves a pseudo $R^2$ value of 0.41. Overall, the fit of the model is moderate, though it is not equally good for all classes. Notably, there is an overprediction of the class ``feature implemented'', which is confused with almost all other classes. When we observe the coefficients based on the marginal effects, we observe the significant relationships reported in Table~\ref{tbl:logit-analysis}. While there are significant relationships for most resolutions, these relationships are only based on the activity within the issues, i.e., the roles of the discussants and the number of comments. This is a very interesting result, as this vice versa means that we do not find any significant relationships between the triggers, issue types, consent interactions, and even labels on GitHub and the resolution of the issues using the multinomial logit model.

\begin{figure}
\centering
\includegraphics[width=\textwidth]{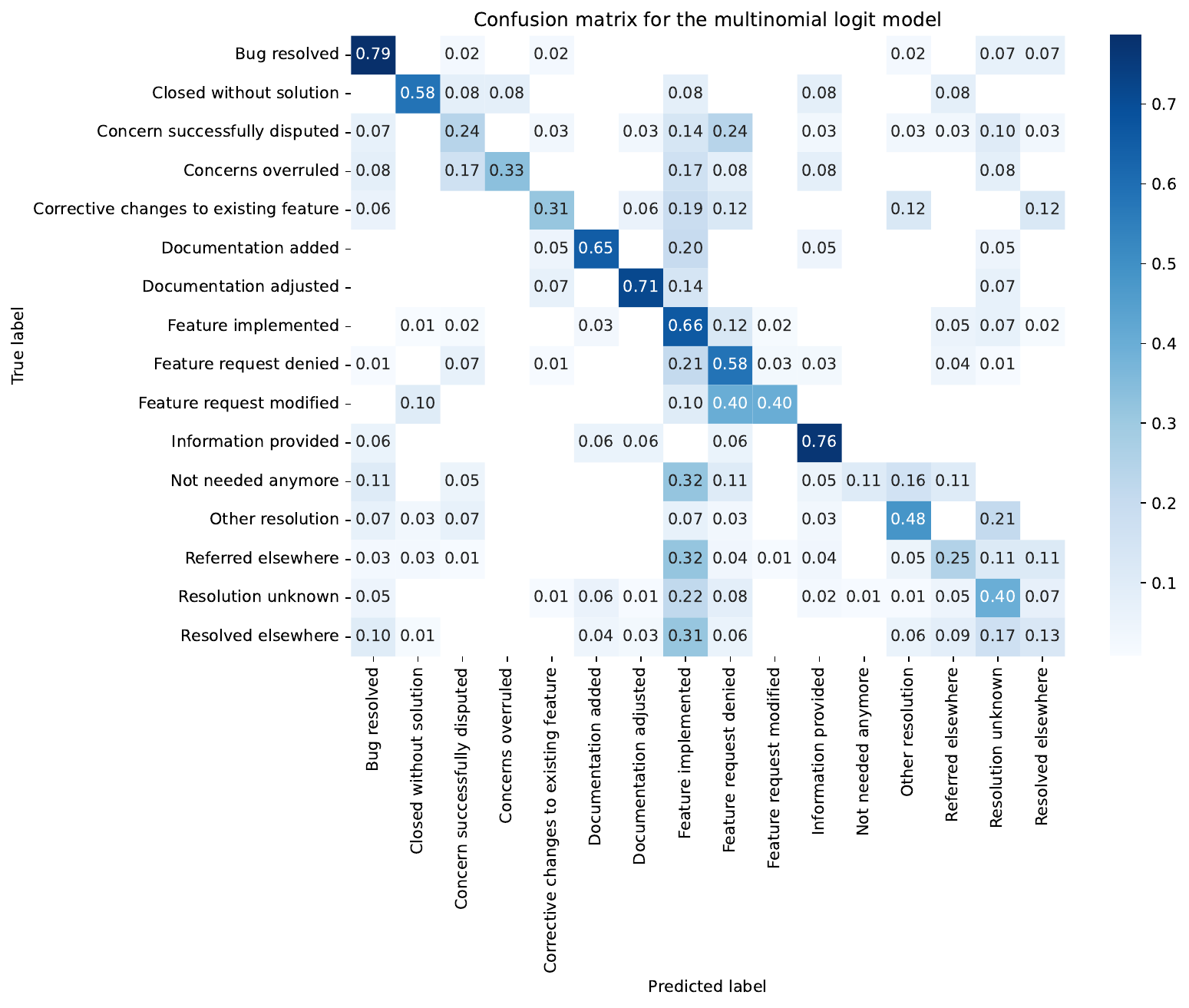}
\caption{Confusion matrix of the multinomial logit model.}
\label{fig:confusion-matrix-logit}
\end{figure}

\begin{table}
\centering
\begin{tabular}{p{3cm}p{8cm}}
\toprule
\textbf{Resolution} & \textbf{Significant relationships} \\
\midrule
Bug resolved & The likelihood decreases when there are more discussants (AME=-48\%, p=0.037) and increases when one-time reporters participate in the discussion (AME=5\%, p=0.036). \\
Closed without solution & The likelihood increases when there are more comments (AME=26\%, p=0.016) and when one-time reporters participate in the discussion (AME=5\%, p=0.026). \\
Concern successfully disputed & The likelihood decreases when there are more discussants (AME=-98\%, p=0.027) and increases when there are more comments (AME=72\%, p=0.031) and when frequent reporters participate in the discussion (AME=6\%, p=0.006). \\
Concern overruled & None. \\
Corrective changes to existing feature & The likelihood increases when there are more comments (AME=52\%, p=0.015). \\
Documentation added & The likelihood increases when one-time reporters participate in the discussion (AME=7\%, p=0.033). \\
Documentation adjusted & The likelihood decreases with more comments within the discussion of an issue (AME=-153\%, p=0.038).\\
Feature implemented & The likelihood increases with more comments (AME=260\%, p$<$0.001) and decreases when there are more discussants (AME=-108\%, p=0.006).\\
Feature request denied & The likelihood decreases when one-time reporters participate in the discussion (AME=-6\%, p=0.034). \\
Feature request modified & None. \\
Information provided & The likelihood decreases when there is no specific trigger for reporting the issue (AME=-3\%, p=0.022).\\
Not needed anymore & None. \\
Other resolution & None. \\
Referred elsewhere & The likelihood decreases with more comments (AME=-254\%, p=0.003) and increases with more discussants (AME=160\%, \rev{p$<$0.001)}.\\
Resolution unknown & The likelihood increases with more discussants (AME=108\%, p=0.021).\\
Resolved elsewhere & None. \\
\bottomrule
\end{tabular}
\caption{Results from the multinomial regression of the resolution types. The table reports the significant average marginal effects, i.e., the average change in probability of the probability of the resolution type if the value of the significant independent variable changes by one.}
\label{tbl:logit-analysis}
\end{table}

\begin{figure}
\centering
\includegraphics[width=\textwidth]{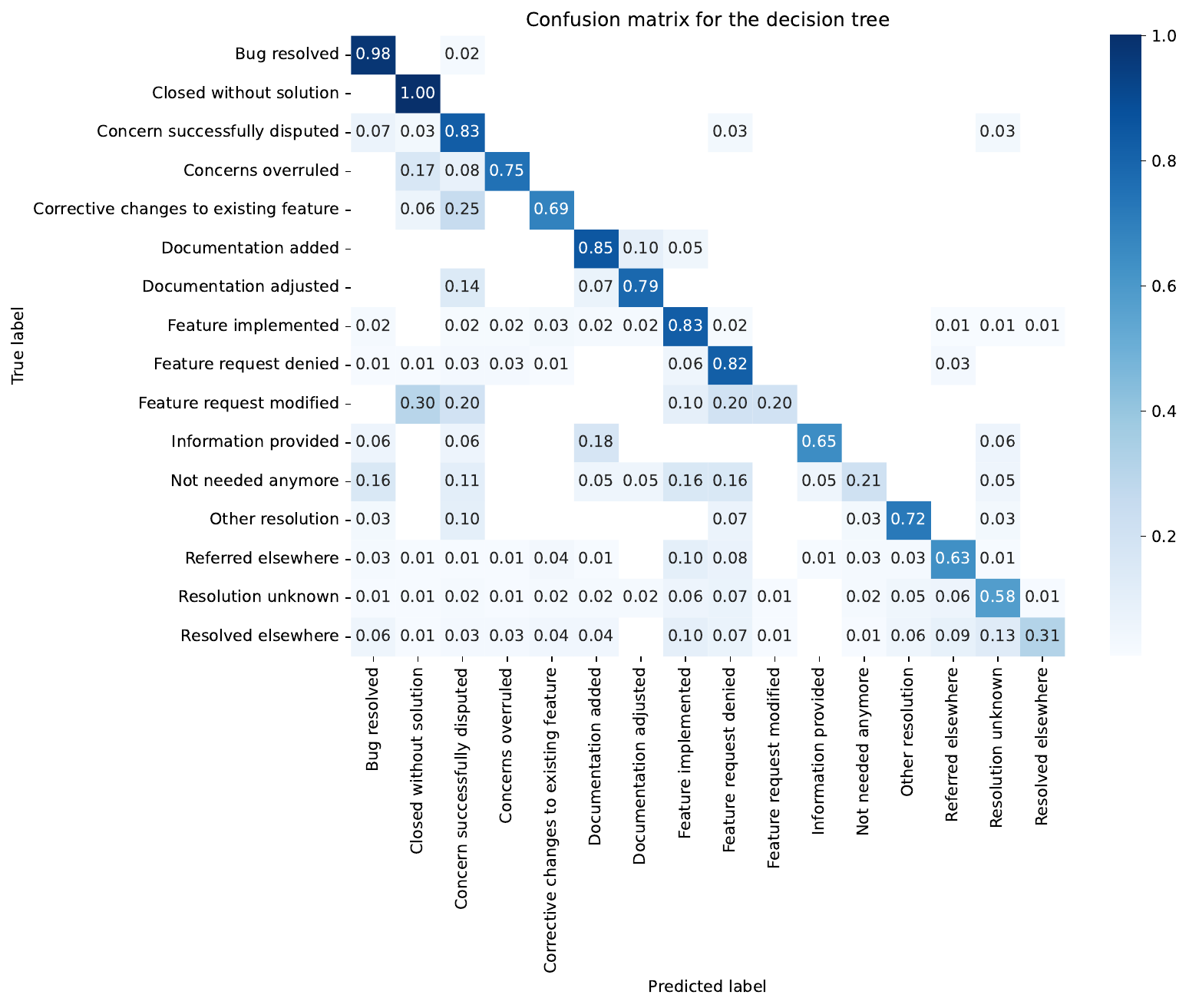}
\caption{Confusion matrix of the decision tree.}
\label{fig:confusion-dt}
\end{figure}

As can be expected, given that we do not restrict the decision training with pruning, the confusion matrix of the decision tree on the training data depicted in Figure~\ref{fig:confusion-dt} shows an overall very strong performance. However, even though the tree is allowed to overfit, not all classes can be accurately identified, which indicates that while for most instances there is a unique combination of feature values, the features sometimes cannot be separated. Notably, the resolutions ``feature request modified'', ``not needed anymore'', and ``resolved elsewhere'' are classified correctly in less than 50\% of the cases. The decision tree we learned had 515 nodes in total, 258 leave nodes, and a maximum depth of 24. We manually analyzed this tree structure from the root to each leaf node. This analysis showed that most of the strong performance indeed results from overfitting, i.e., splitting relatively impure combinations of resolutions again and again, until only 2 instances are remaining. Without such overfitting, the tree almost never finds fairly pure combinations. 

However, this does not mean that the decision tree did not yield interesting insights. Before the tree broke the decision down through overfitting, there were higher level decisions, which in some cases led to a majority of the samples matching the current path of the tree falling into few resolution categories. We observed the following notable patterns: 
\begin{itemize}
    \item The first decision was whether an issue was labeled as bug. Of the 67 issues, for which this was the case, 23 were later assigned the label ``bug resolved'', an additional 12 were resolved elsewhere. When the issue was opened by a frequent reporter, 13 of 24 were labeled as ``bug resolved''. Thus, if issues are labeled as bugs, there is about 50\% chance that they will be resolved. If they come from a frequent reporter this resolution likely happens directly within the project. 
    \item For issues not labeled as bugs, the next decision was whether they were not information requests and whether they were additionally labeled as enhancements. Of the 145 issues matching these criteria, 55 of 145 issues were resolved as ``feature implemented'', another 23 issues were resolved as ``referred elsewhere'' and 19 were ``resolved elsewhere'', while only 17 were resolved as ``feature request denied''. Thus, about two-thirds of not-information-seeking enhancements are resolved. 
    \item For the issues not labeled as bugs, that were not information seeking, but also not labeled as enhancement, it depends on who reports these issues. If the issues come from a frequent committer, the outcome for the 147 issues matching these criteria is similar to having an enhancement label: 44 are resolved as ``feature implemented'', 14 are ``referred elsewhere'', and 19 are ``resolved elsewhere''. Only five such issues were resolved as ``feature request denied''.
    \item In contrast, for the 274 issues where the reporter is not a frequent committer, 48 were resolved as ``feature request denied'' and only 28 were resolved as ``feature implemented''. Within the next layers of the tree, further information regarding the denied feature requests can be gained, i.e., that they are mostly related to feature requests with privacy implications (15 instances) or privacy enhancements (18 instances). 
\end{itemize}

\begin{figure}
\centering
\includegraphics[width=\textwidth]{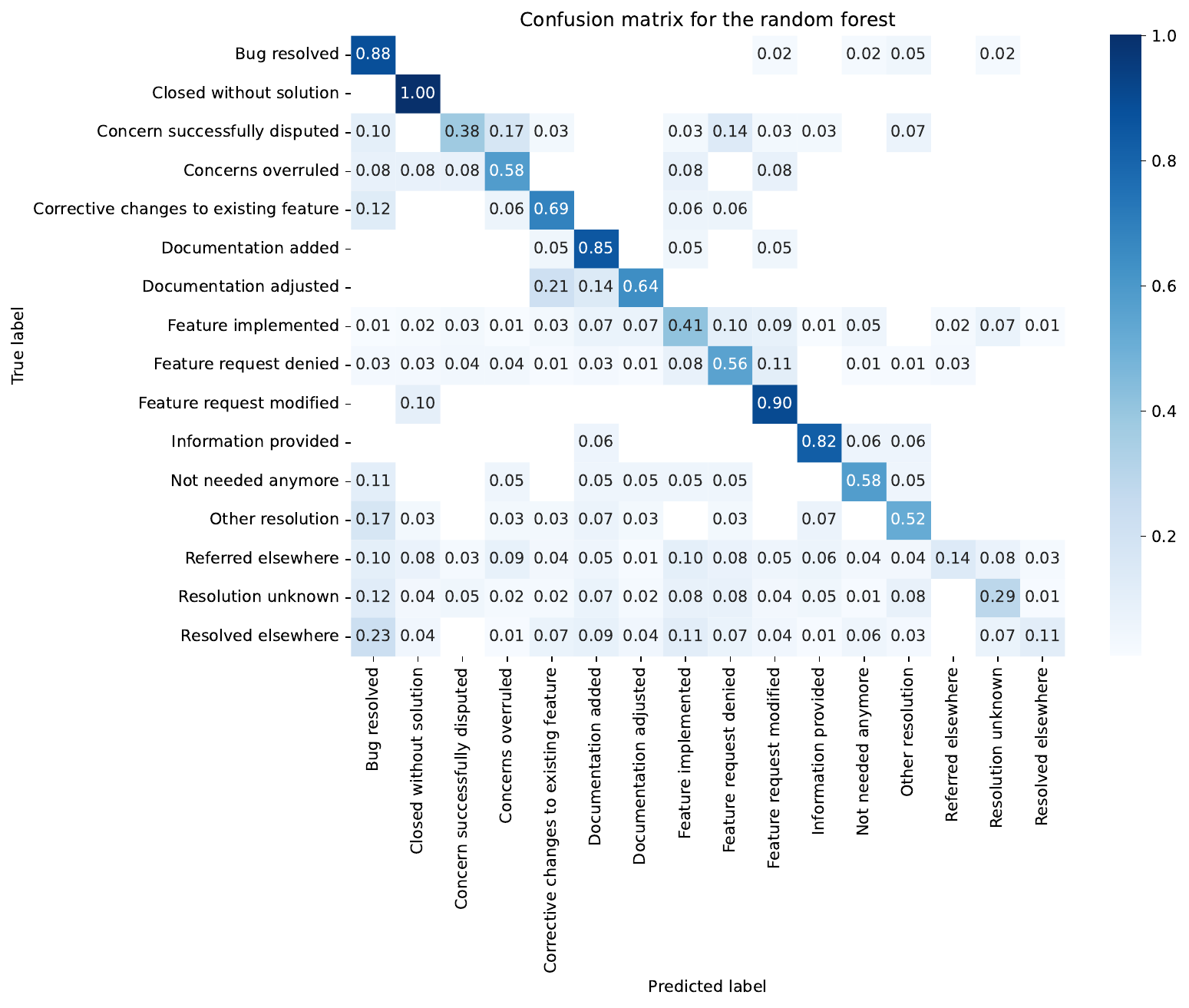}
\caption{Confusion matrix of the random forest.}
\label{fig:confusion-rf}
\end{figure}

The random forest demonstrates that the relationship between the resolutions and our independent variables can be learned to some degree, but not reliably for all classes. Our features do not seem to be well-suited to identify if issues are resolved elsewhere, their resolution is unknown, or whether they were successfully disputed, as for all of these classes less than 40\% of the instances are classified correctly by our random forest. Interestingly, the random forest is also not able to reliably identify if features were implemented, as this is also only correct for 41\% of the instances. The confusion matrix shows that nearly all classes are sometimes, but not often, mixed up with features being implemented (see Figure~\ref{fig:confusion-rf}). This aligns well with what we observed with the decision tree. There, we also found that it was hard for the model to distinguish features being implemented from other classes. However, we note that since we have sixteen classes, even these lower accuracies are a lot better than a random classification, which would only yield an accuracy of about 6\%. Thus, the random forest shows that there is indeed a relationship between our variables. 

\begin{figure}
\centering
\includegraphics[width=\textwidth]{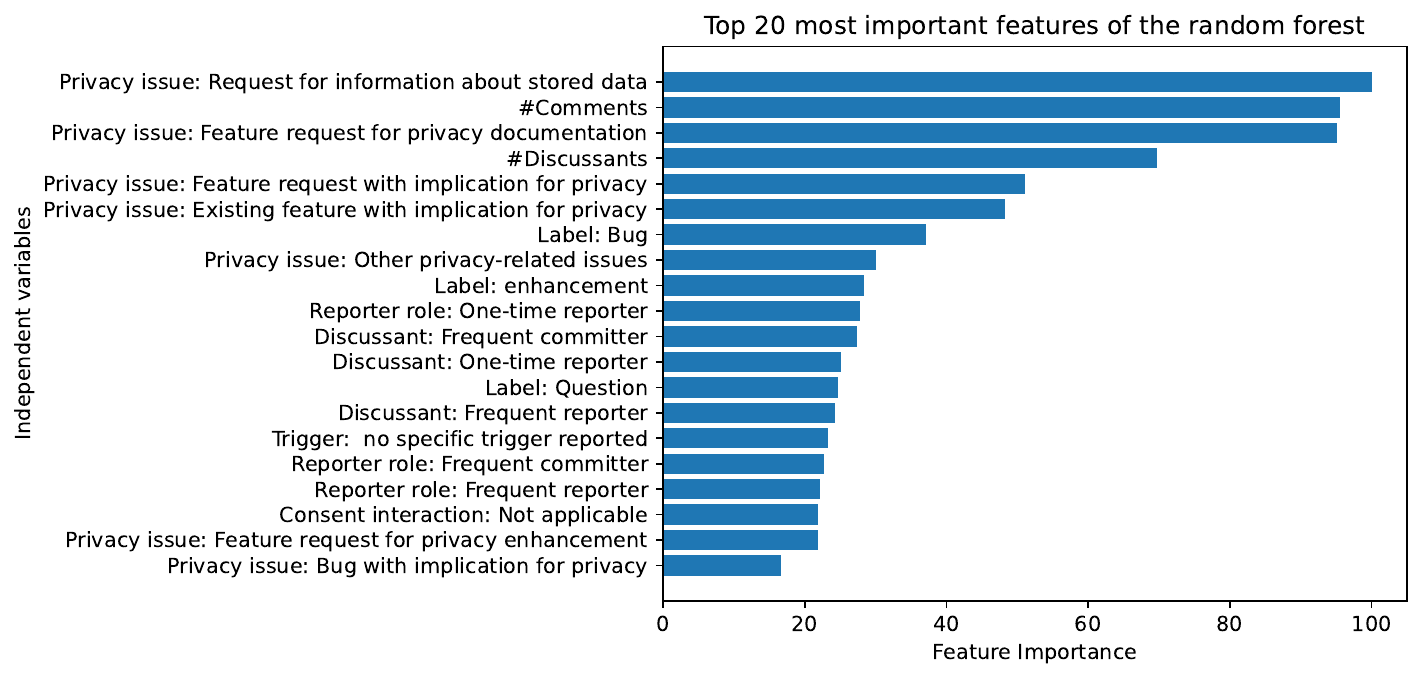}
\caption{Feature importance for the random forest. The feature importance is the ratio of the number of decisions within the random forest, that is based on a feature. We report the feature importance as percentages in relation to the most important feature.}
\label{fig:var-imp-df}
\end{figure}

When we consider the relative importance of the independent variables for the decisions of the random forest (see Figure~\ref{fig:var-imp-df}), three of our variables stand out: whether the issue is a request for information, it is regarding privacy documentation, and the number of comments. Other variables follow with larger gaps: the number of discussants, whether a new or existing feature with privacy implications is discussed, or whether the label is bug.

\summaryboxnew{\rev{The three predictive model show that there is a relationship between our variables and the resolution of privacy issues. Unsurprisingly, activity related variables (e.g., \#Comments) have a strong impact on whether issues are resolved, who is involved (e.g., frequent committers) is also a factor, but relatively weak. Moreover, we note that documentation is frequently updated and that bugs are more likely to be addressed then feature requests.}}

\section{Discussion}
\label{discussion}

As described in Section~\ref{results}, we analyzed 652 GitHub issues that were opened between April 2016 (the year in which the GDPR was adopted) and December 2022. We sampled issues from public GitHub projects that are reported in English and are dealing with personal data or data protection topics. 

By analyzing these issues we aimed to answer the following research questions, for which we will discuss the results in the following. 
\begin{description}
    \item RQ 1: \textit{What kind of issues related to personal data and data protection are discussed on GitHub?}
    \item RQ 2: \textit{How often are those issues related to personal data and data protection reported?}
    \item RQ 3: \textit{Who reports and discusses issues related to personal data and data protection on GitHub?}
    \item RQ 4: \textit{How do developers react to such reported issues?}
\end{description}

\subsection{RQ 1: What kind of issues related to personal data and data protection are discussed on GitHub?}

We were surprised that, in general, discussions on personal data and data protection appear in many diverse contexts. As one may have expected, the majority of issues are opened to enhance users' privacy, e.g. requests for new privacy enhancing features, like anonymizing or deleting personal data in databases. This also includes features for regulation compliance, like disabling third-party data collection\rev{,} or consent interaction, like adding a cookie banner. On the other hand, the second-most type of issues in our data were feature requests that would have negative implications for users' privacy, i.e., features that would expose or collect personal data, like location data or analytics tools. We saw that those requests were mainly denied (26/70), but also saw that at least 15/70 were implemented, while 4 were resolved and 5 were referred elsewhere and are likely to be implemented there.

Furthermore, we found it notable that GitHub is also used to evaluate the compliance of a project, or to request information about stored data. We saw that in most of the cases information, e.g, in form of a documentation, was provided and only a few projects explicitly stated that GitHub is not the right place to request or discuss those issues. This highlights the central role of GitHub issues for communication within a project. 

Although the subject of our study were primarily issues related to personal data and data protection, we also found quite a few security-related issues with the keywords we used. This indicates, as already discussed in previous work~\citep{Balebako.2014, Hadar.2018, Peixoto.2023, Kekulluoglu.2023}, that especially for developers privacy aspects are often intertwined with or superseded by security aspects.

Additionally to our registered study design, we also analyzed whether some\rev{thing} -- and if so, what -- triggered the reporting of an issue. Interestingly, we did not identify any specific trigger for the vast majority of our issues. It seems that any relation to privacy is a trigger itself, so in most of the cases it does not need a regulation or law to start a discussion. This would also be an explanation why \cite{Tahaei.2020} found that laws and regulations are the least common drivers for discussions about privacy-related questions on Stack~Overflow. Nevertheless, we also saw issues where data protection laws, and here mainly GDPR or European privacy laws in general, were cited to add weight to a specific argument.\footnote{e.g., \url{https://github.com/rust-lang/crates.io/issues/955}}

\summaryboxnew{\rev{In summary, we could see that GitHub is a platform where issues on personal data and data protection are discussed in a wide variety of contexts, making GitHub a valuable source for future research to learn how certain topics are discussed within the software development community. Furthermore, we could show that data protection is discussed far more widely within the software development process than in the general public. For example, while the design of cookie banner and obtaining informed consent seem to dominate the public discourse, this is only a small part of the issues related to personal data and data protection reported on GitHub. In contrast on GitHub, concerns about users' privacy are discussed in a much broader context, e.g., related to messenger apps revealing personal information when showing if a person is currently online, or transportation apps requesting precise location data. We could also observe that among GitHub reporters, PII (personal identifiable information) are defined much stricter than, e.g., in the GDPR and, thus, requests for privacy enhancement are in most cases beyond privacy standards that are defined in data protection regulations. This in an interesting finding that should be reflected in the curricula for software developers, who are likely to be encountering such requests later in their careers.}}

\subsection{RQ 2: How often are those issues related to personal data and data protection reported?}

In total, we identified over 12,000 issues from around 5,000 projects that matched at least one of our keywords defined in Table~\ref{tbl:keywords}. When we manually validated if the detected issues indeed match discussions about personal data or data protection, we excluded nearly two-thirds as false positives. Nevertheless, this still means that there is a substantial amount of issues related to data protection being reported on GitHub. Still, compared to the expected number of issues in all projects that meet our inclusion criteria, this number is likely small. While we do not know the exact number, this is just a small fraction of the overall issues on GitHub. As an example, while we ha\rev{d only one} issue from the \rev{``Rust'' project} in our data, the total number of issues for that project alone is over 50,000.\footnote{\url{https://github.com/rust-lang/rust/issues}}

If we look at the trend over time, we see that privacy-related issues are always prevalent, since there has not been a time in the last six years when no issues on personal data or data protection have been reported. Which is also in line with the findings of \cite{Tahaei.2020}, who observed a continuous increase in the number of privacy-related questions on Stack~Overflow between 2008 and 2019. However, within our analysis we were able to see a strong increase in privacy-related issues from the second half of 2017, which peaks in the first quarter of 2018 and then falls sharply again. This is a clear sign that GDPR, which came into effect in May 20\rev{1}8, triggered a lot of issues related to personal data and data protection, even if legal reasons were not always mentioned as the reporting cause.

It might be interesting for future work to investigate with a representative sample if those assumptions can be confirmed by looking at the distribution of the triggers over time, and whether, e.g., GDPR is also the most prevalent trigger for issues on personal data and data protection at the end of 2017 and until the beginning of 2018.

\summaryboxnew{\rev{In summary, we could show that data protection and personal data is actively discussed on GitHub over time. We could also show that the introduction of laws and regulations on data protection is reflected in the software development process. While an in-depth analysis on reasons and correlations was beyond the scope of this study, we can clearly show that regulations have an effect on the software development process and spark activity -- especially shortly before new regulations come into effect. This could be due to some projects seem to wait for other software to decide about GDPR implementation or provide guidelines,\footnote{e.g., \url{https://github.com/awesomemotive/easy-digital-downloads/issues/6216}} which might have caused the sudden peaks in the last quarter of 2017. Future work could provide more fine-granular data on these activities, e.g. by correlating the analysis over time with the resolution or the triggers.}}

\subsection{RQ 3: Who reports and discusses issues related to personal data and data protection on GitHub?}

The issues we analyzed are often reported by individuals who are active in the project independent of the \rev{specific} data protection issues, i.e., frequent reporters and frequent committers. However, there are also many one-time reporters, who request privacy improvements. 

The discussions paint a different picture: the frequent committers play a key role, while frequent reporters also contribute substantially. In contrast, individuals who are only interested in reporting a single issue seldom participate in the subsequent discussion of the issue.

These results somewhat contradict those of~\cite{Bissyande.2013}: While we also see that issues are mainly reported by individuals who are already active in the projects (either as frequent reporters or frequent committers), we also observe a lot of one-time committers. But the minority of one-time committers also contributes to the code or the discussion.

\summaryboxnew{\rev{In summary, our analysis shows that those already involved in a project are the persons being most active in reporting and discussing issues. Since previous studies with software developers have shown that in terms of handling data protection issues, lack of resources~(\cite{Balebako.2014}), lack of knowledge about privacy best practices~(\cite{Balebako.2014, Hadar.2018}), sufficient understanding of data protection concepts~(\cite{Hadar.2018}), and lack of knowledge about privacy preserving techniques~(\cite{Hadar.2018}) are major challenges, which could, e.g., be addressed by encouraging different perspectives from persons outside of the development team. One-time reporter could contribute valuable skills and knowledge, and workload could be better distributed, especially for Open Source projects, where developers are mainly working on new features or bug fixes in there free time without any payment. It might, thus, be worthwhile for future work to investigate why one-time reporters are not committed to ``their'' issues, and how they -- as well as one-time committers -- could be encouraged to contribute to the projects.}}

\subsection{RQ 4: How do developers react to such reported issues?}

\rev{If we consider the trend over time and the sharp increase in issues on personal data and data protection around the time the GDPR came into effect, we see that -- although not loved by everyone\footnote{e.g., \url{https://github.com/Bforartists/Bforartists/issues/195}} --  data protection laws and regulations were effective in sparking discussions about personal data and data protection within the software development community. In our qualitative analysis, we found that some issues mention time pressure, and relate to GDPR becoming effective soon.\footnote{e.g., \url{https://github.com/ibericode/mailchimp-for-wordpress/issues/526}}}

\rev{For those issues that were triggered by GDPR or other privacy laws, the request had mainly been granted. If a request was denied, this often happened when the developers stated privacy concerns -- in case a feature with implications for users' privacy was requested\footnote{e.g., \url{https://github.com/indentlabs/notebook/issues/90}} -- or when they could dispel concerns that the feature has privacy implications.\footnote{e.g., \url{https://github.com/chocolatey-community/chocolatey-packages/issues/588}} If a request was referred elsewhere, this mainly happened because the request was transferred to or merged with another issue, e.g., a more specific issue with concrete tasks.\footnote{e.g., \url{https://github.com/zonemaster/zonemaster-backend/issues/317}}}

Our quantitative analysis revealed that the activity within an issue is the best predictor for its resolution, i.e., the type of data protection issue does not seem to influence the developers' reactions in general. What seems to improve the likelihood of being resolved is when issues are labeled as bugs on GitHub. Consequently, especially when it comes to reporting problems, tagging them as such through a GitHub label can be a strategy for reporters to increase the likelihood of the issue being resolved. Labeling an issue as enhancement can also help. On the other hand, our data also shows that incomplete reporting, where additional information needs to be requested, decreases the likelihood of issues being resolved. 

But, according to our qualitative analysis, the challenges mentioned above rarely influenced the resolutions we observed. We must acknowledge that we could not track internal discussions, e.g., if help was sought by colleagues or external advisors outside of GitHub. But from the comments, in most cases an issue was resolved, either as requested without further discussion,\footnote{e.g., \url{https://github.com/bobintetley/asm3/issues/310}} or because it was already implemented.\footnote{e.g., \url{https://github.com/ellmetha/django-machina/issues/123}} The second most frequent reason for a resolution was that an issue was transferred to a more appropriate issue for the final implementation.\footnote{e.g., \url{https://github.com/openmobilityfoundation/mobility-data-specification/issues/61}}

On the other hand, we noticed that for 22 issues, developers had to be persuaded to make necessary changes to the project.\footnote{see, e.g., \url{https://github.com/netdata/netdata/issues/7366} for a most notable issue.} This reflects the findings of \cite{Sheth.2014}, who found that users' expectations and needs regarding privacy differ from those of developers. Nevertheless, as this was only observed within less than 4\% of our issues, we assume that in the majority of cases, developers meet the requests of the users.

When requests were not met, it was mainly because no resolution could be found. We cannot tell for all issues if this was because no solution exists (yet) or, if developers were not eager to make changes and just stated that no solution exists.\footnote{e.g., \url{https://github.com/mozilla-services/screenshots/issues/2013}}

As \cite{Balebako.2014} proposed, another challenge why privacy requests are not met might be a lack of resources and/or lack of support from management. According to our data, we cannot see that this is a general trend. Inactivity, a lack of interest, or infeasibility could count as a sign of a lack of time or management support. All three reasons appeared in total 44 times in our data, so only for less than 7\% of the issues. 

This percentage might even be over-representative, because we have to take into account that e.g. inactivity was also coded when the person who opened the issue did not further interact.\footnote{e.g., \url{https://github.com/Nexus-Mods/Nexus-Mod-Manager/issues/188}} Another reason why requests were seen out of scope were technical restrictions.\footnote{e.g., \url{https://github.com/elastic/beats/issues/22770}} Also, some requests went stale because they seemed to be way too specific and not useful for more than one user.\footnote{e.g., \url{https://github.com/telegramdesktop/tdesktop/issues/3230}}

A limitation when discussing developer's motivations on the basis of our data is that we were only able to look at public projects. Some of the projects we found rely on volunteers, which sometimes led to issues only being addressed, if developers themselves support the requested feature, if they have the time to implement it, or if there is enough community support.\footnote{e.g.\url{https://github.com/grote/Transportr/issues/263}} Thus, when we coded \rev{``}lack of interest\rev{''}, this might also be a result of \rev{``}lack of time\rev{''}, i.e., when developers applied cost-benefit analyses, even if not mentioned specifically in the comments.

\summaryboxnew{\rev{In summary, we could show that there are many reasons why an issue is finally addressed. In our data we see that most of the developers are eager to provide solutions and in most cases issues are resolved as requested. While we see that lack of resources sometimes delayed responses or providing a final resolution, we do not see this to be a general trend in our data. However, it seems important that issues are reported comprehensively and labeled correctly to increase the likelihood that developers address them. Since a considerable amount of issues was resolved in a different issue or even outside of GitHub, future work could investigate in more detail how issues are resolved that are transferred somewhere else. This could reveal further challenges that arise during the software development process.}}

\subsection{\rev{Future Work}}

\rev{As our main goals were to provide first explorative insights on how topics related to data protection and personal data are discussed on GitHub, and identify possible patterns and correlations, it was out of scope for this study to investigate our findings in detail. Thus, besides what was already mentioned in the summaries above, we would like to hint at further interesting findings that future work could address.}

\rev{At first, w}e observed that for some issues it took the developers quite some time to address the issue. We assume that time could play a role here as well, but we also felt that for some issues discussions and decisions were made outside of GitHub\footnote{See, e.g. \url{https://github.com/dotnet/SqlClient/issues/11} for a feature request that was finally implemented, but where developers' did not communicate well with the community.}. This is most prominent for 48 issues, where we could not define any reason for the given solution. Without insights into the project teams, e.g., interviews with the developers, the concrete reasoning for those issues remains unclear.\footnote{see, e.g., \url{https://github.com/brookhong/Surfingkeys/issues/357} for an issue where no reason for the resolution could be defined} \rev{We also observed that part of the reason why resolutions were delayed was because handling of privacy issues was done outside of the scope of core development, e.g., by involving legal departments\footnote{see, e.g. \url{https://github.com/Automattic/jetpack/issues/10271}} or designers\footnote{see, e.g., \url{https://github.com/brave/brave-ios/issues/1938}}. This shows that we need to better understand the interactions between what developers know about data privacy, features related to this, and the public presence of projects that often publishes notices about data privacy.} 

Furthermore it would be interesting to quantify the time it takes the developers to address an issue either in correlation to the issue type, the number of active discussants or the trigger. Anecdotally, we can report about one issue that was not addressed for about a year, but then the developers recognized that it reports a severe security bug. The issue was then resolved very quickly, but future work might look into possible correlations in more depth.

\rev{Another interesting finding is that the second most reported issue-type are requests for features that would have implications for users' privacy. It would be interesting to further investigate the roles of the persons involved in these issues and who has which position in the discussion. We could, e.g., see that these issues are mainly reported by frequent reporters and that they are mainly denied, but future work could investigate the course of the discussion, the positions that participants take within the discussion, and the way a final resolution is reached and perhaps justified.}

\rev{It was also beyond the scope of this study to investigate emotions in issues related to data protection and personal data. Especially, but not exclusively, for issues were existing features or bugs are reported, or new features are requested that have an impact on users' privacy. It would be interesting to see whether in relation to issues where features are requested that would enhance users' privacy, we see a difference in the tone of the discussions.}

\rev{And, finally, we found that in the vast majority of issues in our sample the reporters or discussants were not openly negative towards data protection regulations. As mentioned above, we did not investigate emotions in detail, so it would be interesting to see if this neutrality is reflected in the population of all issues on data protection and personal data, or if we, by chance, only sampled for the more neutral or positive issues.}

\subsection{Threats to Validity}

We report the threats to the validity of our work following the classification by \cite{Cook1979} suggested for software engineering by \cite{Wohlin2012}. Additionally, we discuss the reliability as suggested by \cite{Runeson2009}. 

\subsubsection{Construct validity}
The construct of our qualitative analysis assumes that the categories we code are meaningful to describe the activity regarding data privacy issues in the wild on GitHub. If our categories were incapable of encoding valuable information about the privacy issues, this information would be lost from our study, possibly altering our results. However, this threat should be negligible. First, we use an inductive coding approach which allows our raters to flexibly incorporate any information they find and which they deem relevant. Second, the raters consider the whole issue, not only aspects directly related to the categories we defined. This actually also led to a deviation in the study protocol, to directly include the trigger for reporting an issue. 

The construct of our quantitative analysis assumes that our variables are suitable to report the salient information regarding our research question and that our measurement devices by reporting counts, correlations, linear modeling, and tree-based approaches are suitable to uncover the desired information pertaining to our research questions. Again, we believe that this threat is negligible: since we used a mixed-methods approach, we also gained an understanding of the issues beyond our variables which indicated that no information was missed. Moreover, the diversity of the analysis approaches means that we consider our research questions from different perspectives, i.e., counts to understand how often something happens, rank correlations to understand direct relationships, and linear modeling and tree-based approaches to understand the relationship to between our variables and the outcome. 

\subsubsection{Internal validity}

Most of the conclusions are direct inferences from our data, and we are not aware of threats to their validity that are not a result of our measurement construct. Notably, the observations from our qualitative analysis align well with the results of our quantitative analysis, supporting the validity of our results. 

\subsubsection{External validity}

While we tried to collect a large and unbiased sample of issues related to personal data and data protection from GitHub, our data collection approach may still introduce some biases. Our criteria for projects exclude small projects with few contributors or general development activity since June 2018. However, we note that activity and contributor-based filtering was identified as a suitable strategy to avoid problems when analyzing data from GitHub~\citep{Kalliamvakou2015}. Moreover, our search-based approach using a list of terms might potentially have missed issues related to personal data and data protection, in case none of the identified terms were mentioned as is, e.g., because we missed terms or due to typos. Due to this, we cannot rule out that we will miss types of issues related to personal data and data protection that are not captured by our search, which would limit the external validity of our study, as our results possibly would not generalize to (open-source) software in general. 

Moreover, since our study is restricted to GitHub as a data source, we cannot generalize our conclusion to reporting of issues related to personal data and data protection in general. For example, users not familiar with software development may not be aware of GitHub and would contact developers in different ways, e.g., social networks or mailing lists. Our study is not suitable to capture such issues, unless the developers would then create an issue on GitHub related to this. In extension to this, we also only capture issues created (and discussed) in English. While we believe that the majority of open source projects that are relevant to our research fall into this category, projects, where discussions happen in different languages, might offer additional insights. Including these, however, is beyond the scope of this paper.

We only consider GDPR (EU), DPA (UK), PA (CAN), and CCPA (CA, USA) as regulations in our keywords {\rev{(see Table~\ref{tbl:keywords})\footnote{Note that in our results we only report on GDPR and CCPA, because none of the other regulations play a role in our analysis, e.g, none of them triggered a discussion or was mentioned in the issue}}}. Including regulations from additional jurisdictions might have broadened our search, but was also beyond the scope of this paper. Nevertheless, we note that within our sample other privacy regulations were not often mentioned, which mitigates this threat. 

\subsubsection{Reliability}
\label{sec:reliability}

Human judgment played a crucial role in our study, as the rating of the privacy issues regarding their type, trigger, consent interaction, and resolution was done manually. As discussed in Section \ref{sec:execution-plan}, we used multiple raters for a subset of the data, to ensure a common agreement. Due to the diverse nature of the possible wording problems, these raters did not reach the desired inter-rater agreement in terms of categories they defined, but when discussing these categories it showed that there was agreement between both raters, when the wording was harmonized. Consequently, while there is always a threat that different human actors come to different results, the high agreement (in content, not wording) between our raters indicates that it is unlikely that the conclusions of our work would change. However, the presentation of the results would almost certainly be different, e.g., by having different codes, sometimes with slightly different groupings (e.g., merging explicit consent and opt-out interactions). \rev{To further mitigate the risk of having a single rater, even after we established that different people essentially rate the same, a third rater checked 10\% of judgments, i.e., 60 issues. Of these, only for one issue there was a minor mistake, where the resolution was coded as ``resolved elsewhere'', whereas the resolution was in fact done in a commit that referenced the issue. In a second case, the trigger was coded as ``European privacy law'' as this was initially mentioned by the reporter when creating the issue. About a year later, the reporter specifically referenced the GPDR, meaning this would have been an alternative code. Still, this does not have an effect on our results, as we consider the similarity between these triggers in our analysis. Thus, while there may be a lingering risk due to a very low number of wrongly coded issues, it is highly unlikely that this has any impact on our conclusions.}

\section{Conclusion}
\label{summary}

In this work, we set out to understand the ``whos'', ``whats'' and ``whys'' of issues related to personal data and data protection in public GitHub projects. Previous studies have investigated different topics or discussions about personal data and data protection in different channels. To the best of our knowledge, no study has yet researched discussions about personal data and data protection on GitHub\rev{, i.e. how these topics are discussed \textit{during} the software development process.}. Furthermore, related work either focused on discussions that were started by developers themselves, or discussions that were more general and not tied to a certain project. Thus, the contributions of our exploratory study are insights into data protection concerns that are reported for open source software, insights about the reporters and discussants of such issues, as well as insights into the (re-)actions of developers. We, furthermore, shed light on the impact of data protection regulations throughout the software development process specifically on GitHub. 

Our study combined qualitative and quantitative analyses. We used open coding to analyze the trigger, type of privacy issue, consent interaction, resolution, and reasons for the resolution of 652 issues related to personal data and data protection. We also used descriptive statistics to answer our research questions, as well as $\chi^2$ tests to determine relationships between the variables, and a multinomial logit model, decision tree, and random forest to predict the resolution of issues.

We calculated that issues on personal data and data protection were reported around 4600 times between April 2016 and December 2022 for the projects we studied. Over time, discussions of issues on personal data and data protection were always prevalent, but we observed a significant increase in reporting from the second half of 2017 until the first quarter of 2018, which is the time the GDPR came into effect. Reporters came both from within the project, but also from the outside, only active for the reporting of a single issue. However, while people from the outside reported issues, they usually did not participate in the subsequent discussion of an issue. 

We also found that discussions on personal data and data protection are very diverse and appear in many different contexts. Most issues are asking for new features that would enhance users' privacy or implementations that would make the project regulation-compliant. Interestingly, only a minority of issues are triggered by a specific cause, but if they are, it is mainly by GDPR or European privacy legislation in general. Only a few issues are solely triggered by CCPA. We also could observe that most of the requested features have been implemented without discussion, which indicates that data protection laws and regulations are effective in starting discussions about personal data and data protection within the software development community.

While this was the first study of this kind, we hope that future studies will continue our work. Future work could, e.g., use a larger sample to support our findings or investigate if our results also generalize for smaller projects or private projects. Furthermore, it would be interesting to get some developers' feedback in the form of, e.g., interviews. We coded the reasons for the resolutions based on the information in the comments, but for some issues, we could not clearly define what blocked them, possibly some discussion and decisions were made outside of GitHub in a manner that we did not identify.


%
%

\section*{Declarations}
\noindent\textbf{Funding} This research is supported by funding from the topic Engineering Secure Systems, topic 46.23.01 Methods for Engineering Secure Systems, of the Helmholtz Association (HGF) and by KASTEL Security Research Labs.

\vspace{5px}

\noindent\textbf{Ethical Approval} Not applicable.

\vspace{5px}

\noindent\textbf{Informed Consent} Not applicable.

\vspace{5px}

\noindent\textbf{Author Contributions} All authors were involved in planning the work and defining the methodology. Lukas Schulte and Steffen Herbold performed the measurements, Anne Hennig and Lukas Schulte conducted the qualitative analysis while Steffen Herbold performed the quantitative analysis. Anne Hennig, Steffen Herbold and Peter Mayer drafted the manuscript and Steffen Herbold designed the figures. All authors discussed the results and contributed to the final manuscript.

\vspace{5px}

\noindent\textbf{Data Availability}  The datasets generated during and/or analyzed during the current study are available in the Zenodo repository, \url{https://doi.org/10.5281/zenodo.10730369}.

\vspace{5px}

\noindent\textbf{Conflict of Interest}  The authors have no competing interests to declare that are relevant to the content of this article.
        
\vspace{5px}
        
\noindent\textbf{Clinical Trial Number} Not applicable.

\FloatBarrier

\bibliographystyle{spbasic}      
\bibliography{literature}   

\appendix

\section{Code Book for Manual Coding}
\label{sec:codebook}

\textbf{Trigger} 
\begin{itemize}
    \item \textit{none}
    \item[]no specific trigger was named
    \item \textit{personal privacy preferences} 
    \item[]if someone says this should be addressed, because of personal preferences or privacy concerns of a (potential) user or a group of user
    \item[] \textit{Example:} ``I wish to hide the visitors IPs for privacy reasons.''
    \item \textit{GDPR}
    \item[]if only GDPR was mentioned as a reason to open or discuss an issue
    \item[] \textit{Example:} ``Since GDPR regulation, anonymization is (sic!) become an important consideration.''
    \item \textit{CCPA}
    \item[]if only CCPA was mentioned as a reason to open or discuss an issue
    \item[] \textit{Example:} ``Decide if CCPA opt-out cascading behavior needs to be updated''
    \item \textit{GDPR and CCPA}
    \item[]if both, GDPR and CCPA are mentioned as reasons to open or discuss an issue
    \item[] \textit{Example:} ``This work prepares COS for proper encryption of events to comply with GDPR/CCPA.''
    \item \textit{GDPR and other privacy laws}
    \item[]if both, GDPR and other regulations or laws (e.g. e-privacy directive or cookie law) are mentioned as reasons to open or discuss an issue
    \item[] \textit{Example:} ``Given the recent changes in laws regarding privacy and the obligation of being conform to the GDPR rules/regulations, I was wondering if InstaPy stores any data of the users'profiles it interacts with.''
    \item \textit{European privacy law}
    \item[]if privacy laws in the EU / in Europe or in a European country (e.g. Germany) in general without further specification (i.e. the specific title of a law or regulation ) are mentioned as reasons to open or discuss an issue
    \item[] \textit{Example:} ``Creates.io does not currently display a privacy notice, and I am concerned that crates.io may be operating in violation of data protection legislation in the EU (and elsewhere).''
    \item \textit{privacy laws in general}
    \item[]if privacy laws in general without further specification (i.e. the specific title of a law or regulation) are mentioned as reasons to open or discuss an issue
    \item[] \textit{Example:} ``Another discussion could be if this practice of disallowing users to disable Admin notifications does not brake (sic!) any current or future privacy related law.'' 
    \item \textit{law in general}
    \item[]if just law is mentioned as reason to open or discuss an issue without any specification (i.e. the specific title of a law or regulation) or reference to privacy specific laws
    \item[] \textit{Example:} ``Cookie banner to inform user over local system cached cookies and performance tracked cookies belongs to many countries law agreement.''
    \item \textit{other trigger}
    \item[] open text 
\end{itemize}

\noindent\textbf{Privacy Issue}
\begin{itemize}
    \item \textit{bug with implication for consent interaction}
    \item[]bug that has implications for consent interaction, e.g., cookie banner not working as intended, accessibility issues, privacy notes not shown, etc.
    \item \textit{bug with implication for privacy}
    \item[]bug that has an implication on privacy, e.g. limits privacy, unnecessary data collection, distinction to ``existing feature with implication for privacy'': the feature does not work as intended, the defective behavior and, thus, implication for privacy was not foreseeable
    \item \textit{cookie banner missing}
    \item[]request for consent interaction specifically related to cookie banner that needs to be added
    \item \textit{compliance evaluation}
    \item[]just documentation or notes on how to comply to current regulations, or an internal evaluation, neither a bug nor a feature request
    \item \textit{existing feature with implication for privacy}
    \item[]feature in an existing tool that has implications for users' privacy (e.g. data leakage, unnecessary data being stored), distinction to ``fr for privacy enhancement'': it is not about adding a new feature but instead changing an existing one, distinction to ``bug with implication for privacy'': the feature that has implications for users' privacy was implemented as intended and it only became clear later that it has implications for privacy (but could have been foreseeable if privacy had been considered by design)
    \item \textit{existing feature with implication for security}
    \item[]feature in an existing tool that has implications for users' security (e.g. unencrypted data storage)
    \item \textit{fr for consent interaction}
    \item[]request for a new feature that deals with user consent, it can be legally motivated (e.g. mandatory agreement to terms of service) or just related to any consent interaction in general (e.g. newsletter subscription)
    \item \textit{fr for privacy documentation}
    \item[]anything related to privacy documentation, e.g. GDPR spec sheet, changes to a privacy notice, translations of privacy notices, etc.
    \item \textit{fr for privacy enhancement}
    \item[]request for a new feature that would enhance users' privacy in an existing tool or a tool in development (e.g. data anonymization)
    \item \textit{fr for regulation compliance}
    \item[]request for a new feature to an existing tool that would make the tool compliant to current (privacy) legislation (e.g. delete user data)
    \item \textit{fr for security enhancement}
    \item[]request for a new feature that would enhance users' security in an existing tool or a tool in development (e.g. adding password checks, protection against phising, etc.)
    \item \textit{fr with implication for privacy}
    \item[]request for a new feature in a tool that would limit users' privacy (e.g. implementing analytics)
    \item \textit{fr with implication for security}
    \item[]request for a new feature in a tool that would limit users' security (e.g. a password that is remembered)
    \item \textit{private information disclosed in discussion}
    \item[]private information (e.g. user data) that are disclosed in a discussion
    \item \textit{request for information}
    \item[]a question that is aimed to gain more information on the privacy or security of a tool
    \item \textit{other privacy issue}
    \item[]everything else or missing codes, please specify in notes
\end{itemize}

\noindent\textbf{Consent Interaction}
\begin{itemize}
    \item \textit{not applicable}
    \item[]the issue is about a topic where no user consent is needed (e.g. issues that deal with privacy, but no consent interaction is expected from the user)
    \item \textit{explicit consent}
    \item[]user consent should be implemented as new feature and it was decided to implement it as explicit consent, it was not specified if opt-in or opt-out will be used
    \item \textit{opt-in}
    \item[]user consent should be implemented as new feature and it was decided to implement it as explicit consent using opt-in
    \item \textit{opt-out}
    \item[]user consent should be implemented as new feature and it was decided to implement it as explicit consent using opt-out
    \item \textit{implicit consent}
    \item[]user consent should be implemented as new feature and it was decided to implement it as implicit consent with no option to opt-out (e.g. only displaying a note that user metrics will be collected)
    \item \textit{consent not properly requested}
    \item[]user consent is not properly requested (e.g. implicit consent is in place while explicit consent would be regulation compliant), also coded if privacy documentation is incorrect and the consent is therefore not valid for the data being collected
    \item \textit{misleading description}
    \item[]user consent is not properly requested due to a misleading description (e.g. if the description of a consent interaction is written in an unclear way)
    \item \textit{no consent required}
    \item[]it was decided that user consent is not necessary (e.g. in a discussion on how to implement a new feature for consent interaction)
    \item \textit{not defined}
    \item[]user consent should be implemented as new feature but it was not defined how consent will be obtained
    \item \textit{other consent interaction}
    \item[]everything else or missing codes, please specify in notes
\end{itemize}

\noindent\textbf{Resolution}
\begin{itemize}
    \item \textit{bug resolved}
    \item[]bug was resolved (code only, if issue was bug)
    \item[] \textit{Example:} ``That works for me. Nice going.''
    \item \textit{closed without solution}
    \item[]there is no solution to this issue, distinction to ``resolution unknown'': it is documented that no solution for this issue exists (at the moment), although the request is valid and privacy concerns will persist
    \item[] \textit{Example:} ``Added the 'unresolved' label although it seems like \#74 (unclosed) should be given priority [...].''
    \item \textit{concern successfully disputed}
    \item[]it was uniformly decided that the privacy concerns do not exist, e.g. the feature has no implication for privacy
    \item[] \textit{Example:} ``Actually it should not be an issue [...]. So we are there creating a hash on the already hashed User Id.''
    \item \textit{concerns overruled}
    \item[]it was unilaterally decided that the privacy concerns do not exist, e.g. if the developers stated that the feature has no implication for privacy even if the user is still uncertain about the argumentation
    \item[] \textit{Example:} ``It's not relevant at all. GDPR is not applying in this case [...].''
    \item \textit{corrective changes to existing feature}
    \item[]the feature that was problematic has been adjusted or changes (code only if issue was feature request)
    \item[] \textit{Example:} ``Yes, this looks good, thanks.''
    \item \textit{documentation added}
    \item[]privacy policies were added as requested (similar to feature implemented, but code only for ``fr for privacy documentation'')
    \item[] \textit{Example:} ``Added documentation for it at the top of the page: [...].''
    \item \textit{documentation adjusted}
    \item[]privacy policies were changed as requested (similar to feature implemented, but code only for ``fr for privacy documentation'')
    \item[] \textit{Example:} ``Specifically, we need to replace: [text] with: [text]'' ``This got done''
    \item \textit{feature implemented}
    \item[]the requested feature was implemented (code only if issue was feature request)
    \item[] \textit{Example:} ``This is now live''
    \item \textit{fr denied}
    \item[]the feature request was denied, e.g. if implementation is out of scope or there are privacy concerns (code only if issue was feature request)
    \item[] \textit{Example:} ``Closing this. Out of scope - at least for now.''
    \item \textit{fr modified}
    \item[]the feature request was modified so that privacy or security implications are no longer an issue
    \item[] \textit{Example:} can only be judged from the context, e.g., \url{https://github.com/odota/core/issues/1250}
    \item \textit{not needed anymore}
    \item[]the requested feature or bug is obsolete and, thus, not needed anymore (e.g. if changes in a beta version were reverted which solved the bug or reverted a feature that had implications for users' privacy)
    \item[] \textit{Example:} ``No idea what has changed but it is working now. Thank you.''
    \item \textit{referred elsewhere}
    \item[]the discussion was moved elsewhere (e.g. to another issue or another application), there is no evidence that the actual issue has been resolved
    \item[] \textit{Example:} ``Thanks for reaching out. Feel free to ping your dedicated account manager or if you don't have any, send email to support{@}adjust.com with this question and our support team will help you with this.''
    \item \textit{resolved elsewhere}
    \item[]the discussion was moved elsewhere (e.g. to another issue or another application), but there is some evidence that the actual issue has been resolved (e.g. commits or merge requests, or a comment in the discussion)
    \item[] \textit{Example:} can best be judged from the context, e.g., \url{https://github.com/decidim/decidim/issues/3320}
    \item \textit{resolution unknown}
    \item[]the issue was closed but there is no evidence that there was a solution or what the solution could have been, distinction to ``closed without solution'': there could have been a solution, but there is no evidence in the issue what has happened
    \item[] \textit{Example:} can best be judged from the context, e.g., \url{https://github.com/brookhong/Surfingkeys/issues/357}
    \item \textit{other resolution}
    \item[]everything else or missing codes, please specify in notes
\end{itemize}

\noindent\textbf{Reason for Resolution (categorized}
\begin{itemize}
    \item \textit{already done}
    \item[]the bug was already resolved or the feature already implemented (e.g. a reason why a feature request was denied)
    \item \textit{discussion rejected}
    \item[]the discussion was limited to collaborators and no further user comments were allowed, mostly the resolution for the issue was unknown
    \item \textit{done as requested}
    \item[]the feature was implemented as requested or the bug was solved without further discussion, sometimes there were further specifications for implementation, but no one was opposed to solving the issue as requested
    \item \textit{done with persuasion}
    \item[]the feature was implemented as requested or the bug was solved but there was some discussion and at least one discussant was opposed to the proposed solution / requested feature, finally the issue was solved as requested
    \item \textit{inactivity}
    \item[]the issue went stale due to inactivity from either the reporter or the disscusants, mostly the resolution was unknown
    \item \textit{intended behavior}
    \item[]the bug was not seen as such because the behavior that was regarded as malfunction was intended
    \item \textit{lack of information}
    \item[]if the developers lack information, e.g. when a bug was filed that could not be reproduced or there were questions on the correct implementation, but no further information that could help to solve the issue were provided by the reporter
    \item \textit{mutual solution found}
    \item[]if, e.g., as a result of the discussion privacy concerns were dispelled or the feature request was modified so that there is no implication for privacy anymore
    \item \textit{no dissenting opinion}
    \item[]the issue was closed because no one disagreed with the reasoning or the resolution
    \item \textit{no further information}
    \item[]the reasons for the resolution of an issue are unclear since no (further) information are provided in the discussion either for a resolution or on the reasons for a resolution
    \item \textit{no implication for privacy}
    \item[]it was decided that the -- either new or existing -- feature has no implication for users's privacy, so either no changes have to be made or a feature can be implemented as requested
    \item \textit{no interest}
    \item[]an issue was closed, because there was obviously no interest in solving the bug or implementing the requested feature
    \item \textit{no solution}
    \item[]an issue was closed if, e.g., there could no solution be found
    \item \textit{not applicable}
    \item[]this code was used when, e.g., personal information were disclosed in a discussion and there was no specific resolution and, thus, no reason provided
    \item \textit{not feasible}
    \item[]if, e.g., a feature request was denied due to lack of time or other ressources
    \item \textit{not responsible}
    \item[]the feature request was denied because the developers are not responsible or have no control over the requested changes
    \item \textit{other solution provided}
    \item[]a workaround or another solution was provided, sometimes also by other users
    \item \textit{privacy concerns}
    \item[]if, e.g., a feature request was denied due to concerns for users' privacy
    \item \textit{resolved by itself}
    \item[]the issue was no longer valid because the behavior did not further occur without anything being changed
    \item \textit{security concerns}
    \item[]if, e.g., a feature request was denied due to concerns for users' security
    \item \textit{transferred to or resolved by a more specific issue}
    \item[]if the issue was, e.g., split up in specific tasks that were resolved in other issues, or if a discussion was merged with another issue
\end{itemize}

\newpage

\section{Mapping of labels from GitHub issues to a unified set of labels}
\label{sec:label-map}

\begin{longtable}{p{2.2cm}|p{8.8cm}}
\caption{Mapping of issue labels used on GitHub to a common set of labels} \\
\toprule
\textbf{Common label} & \textbf{GitHub labels} \\
\midrule
\endfirsthead

\caption{(continued from previous page)}\\
\toprule
\textbf{Common label} & \textbf{GitHub labels} \\
\midrule
\endhead

\toprule
\multicolumn{2}{c}{(continued on next page)}\\
\endfoot
\bottomrule
\endlastfoot

other & 2.2.0, 6.2.0, ?? days, [master], [S] Easy, [Type] Good First Bug, 1. to develop, 1.7.8.6, 13.0, 3rd party compatibility, A11y, AC, administrative, ads, Affected/5.4.0, Affected/5.5.0-Alpha, Affected/5.5.0-Beta, Affected/5.7.0, affects/v4, all users, area/compliance, A-Redaction, BankID, BBC, BC-BREAK, beginner job ??, BO, bounty-m, BPS, CELA, claimed, code quality, compliance, COMRADES, content, CreateAzMonWIFeb2021, Customer Request, Data Science, de, debt, design, difficulty/low, Difficulty: Medium, Doc : Needs Translation, dunno, ed benefits, E-easy, E-mentor, Epic, epic, EPIC, epic 1, FixTheFlows, Github Import, good first issue, google, has:pr, help needed, help wanted, housekeeping, impact/medium, Implement, import, inrupt-sprint, INTENT TO IMPLEMENT, Intent to implement, Issue: Format is not valid, Issue: Format is valid, javascript, kind/epic, last-call, launch, legal, meta, model changes, MVP, need-to-slice, No Code Attached Yet, Notifications, not-in-changelog, Observed, OE, Order, Orphaned, Outdated, papercut, persons, pinned, planned, popular, potential breaking change, pre-cr-p3, PRO, Project, Release : 5.0.0, Release Line: 2.1, release-notes/include, reported, Roadmap, s2s, size/M, size:M, spike, student, suggestion, support, tech-debt, technical-work, Telework Agreement, Temporary Fix Available, testplan-item, tool, unresolved ??, upcoming feature, user, User feedback, User Report, v1, V2, v2.0, v3.x, WUM, Z-REQ-PSM-IA-3.9, Z-REQ-PSM-IA-4.2, Z-REQ-PSM-II-2.2, Z-REQ-PSM-II-2.3, Z-REQ-PSM-II-2.4, Z-REQ-PSM-IU-2.2, Z-TravisR \\

component & (mostly) back-end, :Data Management/Ingest Node, ?? L10N, ?? Scope : Project, [C] Backend, 2. Collectors, 4. Storage, a: stats, Accessibility, accounts, adapter support, A-frontend, Analytics, api, api issue, api: firestore, A-Push, area/component, area/UI-UX, A-Read-Receipts, Area-Input, A-Room-Settings, A-Timeline, A-User-Settings, Azure-Monitor/svc, backend, browser-extension, cat: mod QAGAME, client, cognitive-services/svc, communications, component/NVDA-GUI, copywriting, CoreOS, council-pallet, CSS/SCSS, Customizer, E3, EEG, EmbeddedTemplate, extensions, Extra Sidebar Widgets, firefox, form-builder, frontend, Google Analytics, GrandID, graphql, HTML, i18n, icebox, Infra, interface: Java API, internals, Invoices, ISTS, journal, KFI, l10n, login, metrics, module: cache, module: core, Module: Form Builder, Module: Form Runner, module:client/core, performance, Permissions, platform:mobile, powerbi/svc, powerbi-desktop/subsvc, Product-PowerToys Run, Product-Terminal, Provider, provider/aws, Reader, Redesign, runtime, S3, scope: DAM, Server Side, shared-components, Sharing, small, SSD, telemetry, Theme: Twitter, Topwatchers, tracking, Tracking-External, Type: Ad-Blocking, UI, User voice - VEC, UserVoice, UX/UI, vaadin-app, vscode-website, web, web ui, website, Webviews, WG: analytics, windows, Work: Front-end, ws-articles, ws-media, x:db2, Z-FTUE-Notifications, Z-Mozilla, zola \\

enhancement & :bulb: Enhancement, :bulb: feature,  Feature,  Type: Feature, [Type] Enhancement, [Type] Feature Request, $>$enhancement, add, addition: feature, enhancement, Enhancement, enhancement , enhancement: accepted, Enhancement:1:Minor, feature, Feature, feature request, Feature request, feature request ??, feature/shields, feature: app settings, feature: WebRTC, Feature:Logins, Feature:Onboarding, Feature:OpenInApp, Feature:Privacy\&Security, Feature:PrivateBrowsing, Feature:Search, Feature:Settings, Feature:Sharing, Feature:Tabs, Feature:Telemetry, feature-request, Idea-Enhancement, improvement, kind/feature, new feature, proposal, request: feature, story, Support data suggestion, t: feature, T:feature-request, T-Enhancement, type/enhancement, Type: Enhancement, type: enhancement, Type: Feature, type: feature, type: feature / request, Type: Feature Request, type: feature request, type: new feature, Type: RFE, type:Enhancement, Type:Enhancement, type-enhancement, user story \\

discussion & :bulb: idea, discuss, Discuss, discussion, Discussion, feedback: discussion, Idea, needs discussion, needs: discussion, needs-discussion, Type: Discussion, Type: Idea, type: question or discussion, type:discussion \\

bug & ?? bug, [Type] Bug, 01 type: bug, A BUG, bug, Bug, bug: pending, Bug :bug:, bug :bug:, bug report, C-bug ??, Incompatibility, Issue-Bug, T-Defect, Type/Bug, type: bug, Type: Bug \\

status\_addressed & ?? Status: Ready for Next Release, 4. to release, answered, Close Request, closed/stale, done, eng:ready, fixed, Fixed in 2.1.x, Fixed in 2.2.x, Fixed in 2.3.x, Fixed in 2.4.x, frozen-due-to-age, lifecycle/frozen, Merged, Resolution/Fixed, solved, status:resolved-locked \\

status\_waiting & ?? waiting, [Status] Queued, blocked, blocker, Blocking Next Release, consider soon, hold, inactive?, inactive, on hold / maybe wont, passed first triage, PR Needed, Stale, stale, stale?, status: inactive, Status: Stale, wait for response, waiting, Waiting for user feedback, waiting-response, X-Blocked \\

duplicate & [Closed] Duplicate, close/duplicate, closed/duplicate, duplicate \\

privacy & [Focus] Privacy, [Status] Needs GDPR Review, anonymization, anonymous-consent, c: Privacy, Compliance/GDPR, contract: GDPR, cookie banner, cookies, data privacy, feature/cookies, gdpr, GDPR, privacy, Privacy, Privacy/GDPR, privacy-and-security, team: privacy \\

severity\_low & [P] Minor, low, Minor, P3, P3: Normal, priority.low, priority/low, priority/P3, priority/P4, priority: low, Severity/Minor \\

severity\_high & [Pri] High, [Priority: HIGH], [priority] high, Critical, curated: most wanted, hi prio, high priority, High priority, high-priority, importance/high, important, Important, Major, P1, P1 - Immediate, P1: High Priority, P1: Launch blocker, pri1, Prio 1, Priority 1, Priority/High, Priority/Highest, Priority: High, priority: high, priority-P1, Severity/Blocker, Severity/Critical, Severity: S1, S-Major \\

severity\_medium & [Pri] Medium, [Pri] Normal, P2, P2 - Normal, P-medium, Pri2, Priority - Medium, Priority-2, severity: moderate \\

status\_working & [Status] Accepted, acknowledged, assigned-to-author, In progress, In Progress, in progress, Issue: Ready for Work, need info, Need Information, need input, Needs assignee action, needs attention, needs info, Needs refinement, needs:legal, needs:product, needs-design, Needs-Repro, Refinement Needed, RFC, status: design required, status: needs feedback, Status: Needs Thought, status: Needs-definition, verified, Verified-Int, WIP, work in progress, X-Needs-Design \\

invalid & [Status] Invalid, closed/invalid, Invalid, No change required, non issue, wontfix, worksforme \\

task & 0 - task, Task, task, Type -Task \\

status\_triage & 0. Needs triage, content:new, investigation, Issue: ready for confirmation, needs triage, needs:triage, Triage: Dev.Experience, Triaged, triaged, unconfirmed \\

status\_qa & 3 - to review, awaiting-review, In Review, MR available, PR exists, qa, QA Pass - iPad, QA Pass - iPhone, QA/No, QA/Yes, qa-triaged, Requires testing/validation, Review: Design +1, Review: UX +1, Status:Available, tested, verified by qa \\

documentation & A-Documentation, area/documentation, area: docs-infra, Changelog: Needs Doc, Doc : Needs Doc, doc-bug, doc-enhancement, docs, Documentation, documentation, Issue-Docs, Legal Docs :scroll:, Type/Docs \\

team & Area: Security, area-security, content team, Dev - Sanity Team, Dev - TES/EY, Team:Data Management, Team:Services, teams-developer-support \\

policy & feature policy, policy, Policy \\

question & question, Question, Quizzes \\

security & security, Security, Security: low \\
\end{longtable}

\newpage

\section{Details about keyword search}

\begin{table}[h]
\caption{Statistics about about matches for our keyword search}
\centering
\begin{tabular}{lrrr}
\toprule
\textbf{Keyword} & \textbf{Issues matched} & \textbf{False positives} & \textbf{FPR} \\
\midrule
tracking & 16 & 16 & 1.000000 \\
data sharing & 71 & 66 & 0.929577 \\
fingerprinting & 105 & 93 & 0.885714 \\
data protection & 41 & 34 & 0.829268 \\
cookie prompt & 79 & 64 & 0.810127 \\
privacy controls & 68 & 53 & 0.779412 \\
privacy policy & 4 & 3 & 0.750000 \\
cookie notice & 100 & 75 & 0.750000 \\
personally identifiable information & 203 & 146 & 0.719212 \\
privacy act & 67 & 46 & 0.686567 \\
pseudonymization & 31 & 20 & 0.645161 \\
privacy notice & 191 & 117 & 0.612565 \\
personal data & 5 & 3 & 0.600000 \\
cookie banner & 139 & 82 & 0.589928 \\
data breach & 38 & 21 & 0.552632 \\
PII & 35 & 19 & 0.542857 \\
anonymization & 105 & 57 & 0.542857 \\
consent withdrawal & 2 & 1 & 0.500000 \\
privacy issue & 97 & 48 & 0.494845 \\
privacy violation & 49 & 24 & 0.489796 \\
cookie law & 90 & 44 & 0.488889 \\
privacy problem & 46 & 21 & 0.456522 \\
CCPA & 49 & 21 & 0.428571 \\
privacy settings & 5 & 2 & 0.400000 \\
privacy law & 56 & 21 & 0.375000 \\
right to be forgotten & 46 & 16 & 0.347826 \\
privacy breach & 46 & 14 & 0.304348 \\
GDPR & 84 & 21 & 0.250000 \\
data privacy & 1 & 0 & 0.000000 \\
ePrivacy Directive & 6 & 0 & 0.000000 \\
\bottomrule
\end{tabular}
\end{table}

\section{Variables with low support dropped from the regression analysis}
\label{sec:dropped-variables}

The following variables were non-zero for less than or equal to 25 instances and not considered for the multinomial regression: ``Label: invalid'',
``Label: team'',
``Label: task'',
``Label: security'',
``Label: severity\_low'',
``Label: documentation'',
``Label: question'',
``Label: severity\_medium'',
``Label: discussion'',
``Label: duplicate'',
``Label: status\_triage'',
``Label: status\_qa'',
``Label: policy'',
``Label: status\_addressed'',
``Privacy issue: Bug with implication for security'',
``Privacy issue: Existing feature with implication for security'',
``Privacy issue: Feature request for security enhancement'',
``Privacy issue: Feature request with implication for security'',
``Privacy issue: Private information disclosed in discussion'',
``Privacy issue: Request for information about stored data'',
``Consent: interactionConsent not properly requested'',
``Consent: interactionImplicit consent'',
``Consent: interactionMisleading description'',
``Consent: interactionNo consent required'',
``Consent: interactionOpt-in'',
``Consent: interactionOpt-out'',
``Consent: interactionOther consent interaction'',
``Trigger: CCPA'',
``Trigger: European privacy law'',
``Trigger: GDPR and CCPA'',
``Trigger: GDPR and other privacy laws'',
``Trigger: Other laws'',
``Trigger: Other trigger'',
``Trigger: Privacy laws in general'',
``One-time committer''

\end{document}